\documentclass[twocolumn]{aastex63}

\usepackage{amsmath}
\usepackage{mathtools}

\usepackage{xcolor}

\received{April 2, 2021}
\revised{\today}
\submitjournal{ApJL}

\shorttitle{On the influence of solar wind structures on CME magnetic complexity}
\shortauthors{Scolini et al.}

\begin{document}

\title{Evolution of interplanetary coronal mass ejection complexity: a numerical study \\
through a swarm of simulated spacecraft}

\correspondingauthor{Camilla Scolini}
\email{camilla.scolini@unh.edu}

\author[0000-0002-5681-0526]{Camilla Scolini}
\affiliation{Institute for the Study of Earth, Oceans, and Space, University of New Hampshire, Durham, NH, USA}
\affiliation{University Corporation for Atmospheric Research, Boulder, CO, USA}

\author[0000-0002-9276-9487]{Reka M. Winslow}
\affiliation{Institute for the Study of Earth, Oceans, and Space, University of New Hampshire, Durham, NH, USA}

\author[0000-0002-1890-6156]{No\'e Lugaz}
\affiliation{Institute for the Study of Earth, Oceans, and Space, University of New Hampshire, Durham, NH, USA}

\author[0000-0002-1743-0651]{Stefaan Poedts}
\affiliation{Centre for mathematical Plasma Astrophysics, KU Leuven, Leuven, Belgium}
\affiliation{Institute of Physics, University of Maria Curie-Sk{\l}odowska, Lublin, Poland}



\begin{abstract}
In-situ measurements carried out by spacecraft in radial alignment are critical to advance our knowledge on the evolutionary behavior of coronal mass ejections (CMEs) and their magnetic structures during propagation through interplanetary space.\ Yet, the scarcity of radially aligned CME crossings restricts investigations on the evolution of CME magnetic structures to a few case studies, preventing a comprehensive understanding of CME complexity changes during propagation.
In this paper, we perform numerical simulations of CMEs interacting with different solar wind streams using the linear force-free spheromak CME model incorporated into the EUropean Heliospheric FORecasting Information Asset (EUHFORIA) model. The novelty of our approach lies in the investigation of the evolution of CME complexity using a swarm of radially aligned, simulated spacecraft. Our scope is to determine under which conditions, and to what extent, CMEs exhibit variations of their magnetic structure and complexity during propagation, as measured by spacecraft that are radially aligned.\
Results indicate that the interaction with large-scale solar wind structures, and particularly with stream interaction regions, doubles the probability to detect an increase of the CME magnetic complexity between two spacecraft in radial alignment, compared to cases without such interactions.\ This work represents the first attempt to quantify the probability of detecting complexity changes in CME magnetic structures by spacecraft in radial alignment using numerical simulations, and it provides support to the interpretation of multi-point CME observations involving past, current (such as Parker Solar Probe and Solar Orbiter), and future missions.\ 
\end{abstract}



\section{Introduction} 
\label{sec:introduction}

Coronal mass ejections (CMEs) are large-scale eruptions of magnetized plasma from the Sun \citep{Webb2012}.\ 
Their interplanetary counterparts (often termed ICMEs) can cause severe space weather disturbances at Earth and other planets \citep{Zhang2007, Kilpua2017b, Lee2017, Winslow2020} due to their large kinetic and magnetic energies \citep{Tsurutani1988, Farrugia1993}.\ 
Of particular interest for its role in controlling the solar wind-magnetospheric coupling \citep{Dungey1961} is their internal magnetic field, often interpreted as having a ``flux-rope'' structure \citep{Vourlidas2013}.
 
The CME properties at a given target location are the result of a complex chain of events involving the formation of CMEs at the Sun and their propagation through interplanetary space \citep{Manchester2017}.\
While the existence of a relatively extensive database of CME observations between 0.3 and 1~au has enabled statistical studies of the radial evolution of CME properties in the inner heliosphere \citep{Richardson2010, Winslow2015, Good2016, Jian2018, Janvier2019}, less is known about the evolution of individual CMEs with heliocentric distance in response to external perturbations \citep{Good2019, Vrsnak2019, Salman2020}.\ 

Observational and modeling studies have shown that during their propagation, CMEs undergo a number of changes, particularly as a consequence of the interaction with high-speed streams (HSSs), stream/corotating interaction regions (SIRs/CIRs), the heliospheric current/plasma sheet (HCS/HPS), or other CMEs \citep[e.g.][]{Odstrcil1999a, Manchester2004, Jacobs2005, Winslow2016, Winslow2021, Lugaz2017}. However, the typical evolutionary behavior of interplanetary CMEs is still open for debate: previous studies \citep[e.g.][]{Good2015, Good2018, Good2019, Winslow2016, Winslow2021, Davies2020, Davies2021, Lugaz2020, Lugaz2020b} have showcased a wide span of evolutionary behaviors, ranging from essentially self-similar to strongly non-ideal. Furthermore, the disentanglement of evolutionary effects from the intrinsic spatial variability of CME structures \citep{Lugaz2018} requires spacecraft observations in near-radial alignment (within less than $5^\circ-10^\circ$ of longitudinal separation), which are currently only available for a very limited set of events.

The varying results obtained from previous studies raise the question whether changes in CME structures are an inherent consequence of their interplanetary propagation, or whether they develop as a consequence of interactions with other structures. Ultimately, investigating the evolution of CMEs in response to the interaction with various interplanetary structures requires a multitude of multi-point observations of individual events at different heliocentric distances. Given the lack of such an extensive data set to date, we simulate it using a numerical model and thousands of radially aligned virtual spacecraft \citep[previous efforts include, e.g.,][]{Al-Haddad2019}. {We aim} to answer the following two questions:\
(i) What is the probability for an individual CME to exhibit different magnetic structure types, and increase its complexity, between two radially aligned spacecraft?\
(ii) How does this probability depend on the presence of corotating solar wind structures in the CME propagation space?

This Letter is structured as follows. In Section~\ref{sec:methods}, we introduce the methods and numerical tools used to investigate CME complexity and its changes with heliocentric distance. In Section~\ref{sec:results}, we present and discuss the results of our analysis, and we conclude in Section~\ref{sec:conclusions}. The Appendix contains further details on our methods, including about the spheromak magnetic structure and the CME identification algorithm implemented.\

\section{Methods}
\label{sec:methods}

\subsection{Modeling set-up}
\label{subsec:methods_euhforia}

We perform three-dimensional (3D) magnetohydrodynamical simulations of the inner heliosphere using the EUropean Heliospheric FORecasting Information Asset \citep[EUHFORIA;][]{Pomoell2018} model.
Our simulation domain covers heliocentric distances ($r$) between 0.1 and 2~au, $\pm 80^\circ$ in the latitudinal ($\theta$) direction, and $\pm 180^\circ$ in the longitudinal ($\phi$) direction, employing a uniform grid composed of 
$512(r) \times 80 (\theta) \times 180 (\phi)$ cells.\ 

\begin{figure*}
\centering
{\includegraphics[width=\hsize]{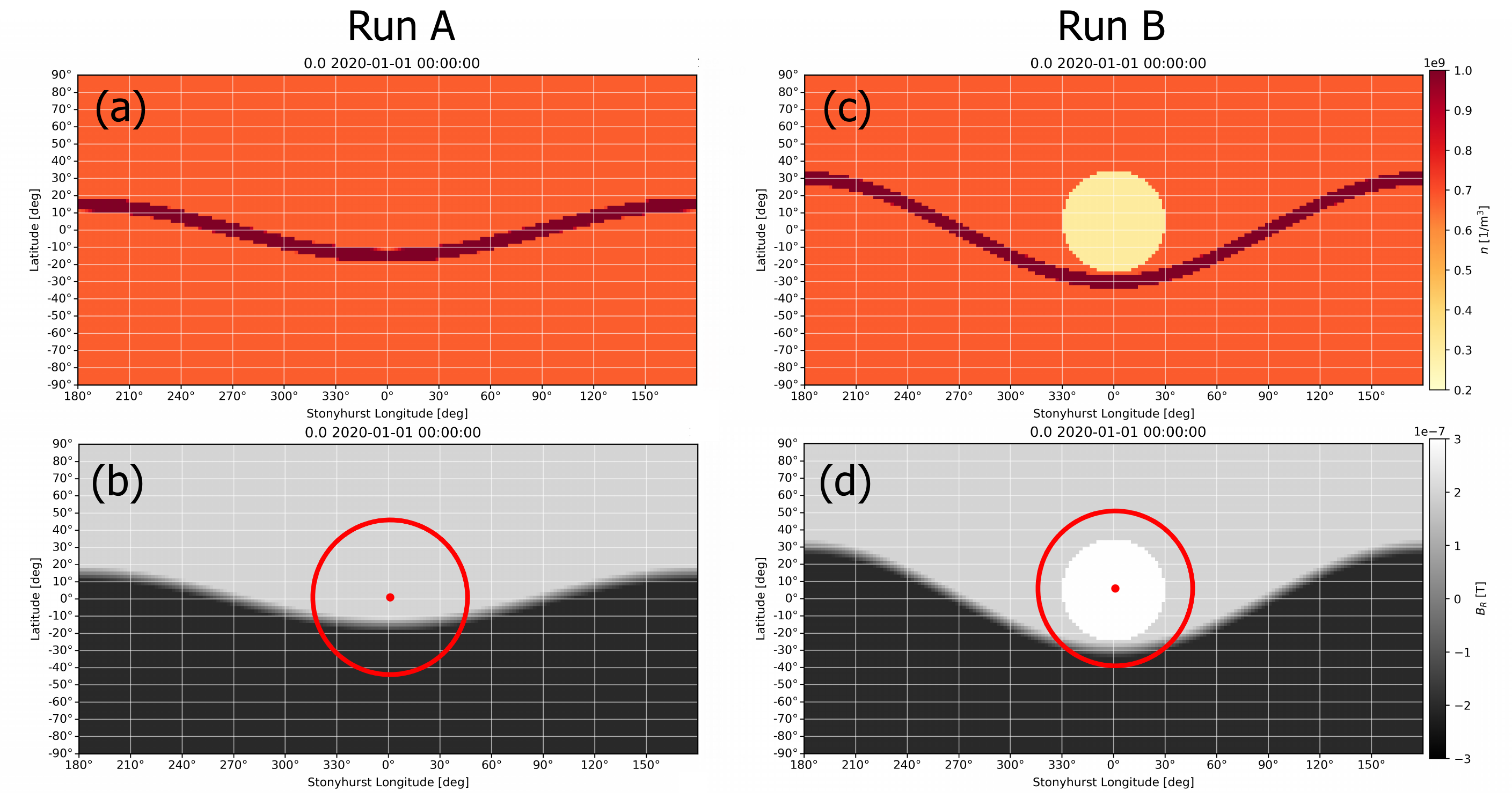}}
\caption{
Solar wind parameters at 0.1~au used as inner boundary conditions in EUHFORIA runs~A (left) and B (right).
(a), (c): number density; 
(b), (d): radial magnetic field, with CME initial direction and cross-section indicated in red.
} 
\label{fig:solar_wind}
\end{figure*}

To evaluate the interaction of CMEs with different solar wind structures, 
we simulate two background solar wind configurations (Figure~\ref{fig:solar_wind}):
the first one (hereafter ``run~A'') {includes a low-inclination HCS/HPS reaching up to $\theta = \pm 15^\circ$, and is} characterized by a uniform solar wind of speed $450$~km~s$^{-1}$ {\citep[intermediate between slow and fast solar wind;][]{Cranmer2017} everywhere else.}
The second one (hereafter ``run~B'') differs from the one above by a HCS/HPS reaching up to $\theta = \pm 30^\circ$, and by the inclusion of a HSS with circular cross-section of {half-}width {(${\omega/2}$)} equal to $30^\circ$ and {radial} speed {(${v_\mathrm{R}}$)} equal to $675$~km~s$^{-1}$, located just above the HCS/HPS at longitude $\phi=0^\circ$.\ 
In both cases, the HPS meridional profile is parametrized using the description in \citet{Odstrcil1996}{, which results in a solar wind speed as low as 300~km~s$^{-1}$ near the HCS}.\ 
{At low latitudes, these idealized configurations mimic the solar wind originating from an equatorial streamer belt (in both runs) and a coronal hole (in run~B). However, they do not include any latitudinal dependence, and are in this respect different from the latitudinal profile observed in the real solar wind \citep{McComas2008}.} 
{This choice has been made} to ensure full control over the CME propagation and interaction with solar wind structures, and the comparability between different runs.\

In both runs, CMEs are initialized at 0.1~au and modeled using the linear force-free spheromak model \citep{Verbeke2019}.\
The following initial parameters are used:
radial speed equal to $800$~km~s$^{-1}$;\
initial half-width of $45^\circ$;\
positive chirality {(${H=+1}$)} with axial tilt {(${\gamma}$)} of $90^\circ$ {with respect to the northward direction} \citep[corresponding to a SWN flux-rope type; see][]{Bothmer1998};\
toroidal magnetic flux {(${\varphi_t}$)} equal to $10^{14}$~Wb (corresponding to a magnetic field strength of $\sim 25$~nT at 1~au).\
Because of the pressure imbalance between the CME and the surrounding solar wind upon insertion in the heliospheric domain \citep[leading to an expansion of the CME structure, as shown by][]{Scolini2019, Scolini2020A&A}, 
the effective initial CME speed is $\sim 1100$~km~s$^{-1}$, which results in a fast CME that drives an interplanetary shock and sheath, as discussed in Section~\ref{sec:results}.\
{Such a combination of initial parameters is representative of those of a typical fast CME with a reconnected flux of the order of $10^{14}$~Wb \citep[][]{Pal2018}.}
The CME initial direction is chosen to reproduce {two end-member scenarios of} interaction with different solar wind structures (shown in panels (b) and (d) in Figure~\ref{fig:solar_wind}):
in run~A, the CME is inserted across the HCS/HPS at $(\theta, \phi) = (0^\circ, 0^\circ)$;
in run~B, the CME is inserted across the HSS at $(\theta, \phi) = (5^\circ, 0^\circ)${, in a configuration similar to that of CMEs originated from ``anemone'' active regions \citep[e.g.][]{Lugaz2011, Sharma2020}.}
The CME insertion time is arbitrarily set on January 1, 2020 at 00:00~UT.\
{A summary of the solar wind and CME parameters at 0.1~au used in this work’s EUHFORIA simulations is provided in Table~\ref{tab:setup}.}

\begin{table}
\begin{center}
 \begin{tabular}{ll} 
 \hline\hline
 \multicolumn{2}{c}{Solar wind parameters -- Runs A and B} \\
 \hline
 $v_\mathrm{R,sw} = 450$~km~s$^{-1}$    & $n_\mathrm{sw} = 675$~cm$^{-3}$ \\
 $B_\mathrm{R,sw} = 200$~nT             & $T_\mathrm{sw} = 3.5 \times 10^5$~K \\ 
 \hline\hline
 \multicolumn{2}{c}{HSS parameters -- Run B only} \\
 \hline
 $v_\mathrm{R,HSS} = 675$~km~s$^{-1}$    & $n_\mathrm{HSS} = 300$~cm$^{-3}$ \\
 $B_\mathrm{R,HSS} = 300$~nT             & $T_\mathrm{HSS} = 0.8 \times 10^{6}$~K \\ 
 $\theta_\mathrm{HSS} = 5^\circ$, $\phi_\mathrm{HSS} = 0^\circ$ & $\omega_\mathrm{HSS}/2 = 30^\circ$ \\
 \hline \hline
 \multicolumn{2}{c}{HCS/HPS parameters -- Run A [B]} \\
 \hline
 \multicolumn{2}{l}{$B_\mathrm{R, HPS}(\theta) = B_\mathrm{R,HSS} \frac{v_\mathrm{R,sw}}{v_\mathrm{R,HSS}} \frac{\theta+\theta_\mathrm{HCS}(\phi)}{\Delta \theta_\mathrm{HPS}}$} \\
 $\theta_\mathrm{HCS}(\phi) = \theta_\mathrm{max} \cos \phi$ & $\theta_\mathrm{max} = 15^\circ [30^\circ]$ \\
 $\Delta \theta_\mathrm{HPS}/2 = 4^\circ$ & $(n v_\mathrm{R}^2)_\mathrm{HPS} = 1.4 \times 10^{20}$~m$^{-1}$~s$^{-2}$ \\
 $P_\mathrm{tot, HPS} = 19.2$~nPa & $T_\mathrm{HPS} = 3.5 \times 10^5$~K \\
 \hline \hline
 \multicolumn{2}{c}{CME parameters -- Run A [B]} \\
 \hline
 $\theta_\mathrm{CME} = 0^\circ [5^\circ]$,  $\phi_\mathrm{CME} = 0^\circ$ & $\omega_\mathrm{CME}/2 = 45^\circ$\\
 $v_\mathrm{R, CME} = 800$~km~s$^{-1}$              & $\rho_\mathrm{CME} = 10^{-18}$~kg~m$^{-3}$ \\
 $\varphi_\mathrm{t, CME} = 10^{14}$~Wb    & $T_\mathrm{CME} = 0.8 \times 10^{6}$~K \\
 $H_\mathrm{CME}=+1$                             & $\gamma_\mathrm{CME} = 90^\circ$ \\
 \hline
 \end{tabular}
 \end{center}
 \caption{\label{tab:setup} Summary of the solar wind and CME parameters at 0.1~au used in this work's EUHFORIA simulations.
 $v_R$ -- radial speed; 
 $B_R$ -- radial magnetic field;
 $n$ -- number density; 
 $\rho$ -- mass density; 
 $T$ -- temperature;
 $P_\mathrm{tot}$ -- total (thermal+magnetic) pressure;
 $\theta$ -- latitude; 
 $\phi$ -- longitude; 
 $\omega/2$ -- half-width;
 $\Delta \theta/2$ -- latitudinal half-width;
 $\varphi_t$ -- toroidal magnetic field;
 $H$ -- chirality;
 $\gamma$ -- axial tilt.
 }
\end{table}

In each simulation, we place virtual spacecraft spanning $\pm 90^\circ$ in longitude from the CME initial direction, and covering the full range of latitudes in the domain.\ The virtual spacecraft are equally distributed with longitudinal and latitudinal separations of $5^\circ$, and are uniformly distributed in the radial direction between 0.11 and 1.61~au (i.e.\ from the model inner boundary to the orbit of Mars) with a 0.1~au separation.\ Overall, a swarm of 18944 virtual spacecraft (1184 per heliocentric distance) is placed in the model domain in each simulation.

\subsection{Identification and classification of CME structures}
\label{subsec:methods_classification}

At each virtual spacecraft, the start of the CME-driven perturbation (i.e.\ a shock-like discontinuity) is determined through our algorithm by scanning the radial speed, density, and magnetic field time series forward in time and applying conditions similar to those typically used to detect fast-forward interplanetary shocks at 1~au \citep[e.g.][]{Vorotnikov2008, Kilpua2015}.\ The detailed identification criteria are presented in Appendix~\ref{subsec:appendix_algorithms_shock}.\

At locations where a CME-driven perturbation is detected, time series are scanned in order to assess whether there is a magnetic ejecta (ME) after the shock-like discontinuity.\ The exact criteria used to determine the ME start and end times vary with heliocentric distance and for different solar wind regimes (as discussed in Appendix~\ref{subsec:appendix_algorithms_ejecta}), but are overall based on two typical characteristics of MEs \citep{Burlaga1981, Kilpua2017}: an enhanced magnetic field strength, and a low plasma $\beta$ compared to the surrounding solar wind.\

After having identified the nominal boundaries of the in situ CME substructures (i.e.\ sheath, ME) at each virtual spacecraft, we classify the ME signature using a classification scheme inspired by \citet{Nieves2019} and based on the amount of rotation of the magnetic field components, i.e.\ $B_R$, $B_T$, and $B_N$, in the local radial-tangential-normal (RTN) coordinate system.
This analysis provides information on the ME structure that is later used in Section~\ref{sec:results} to investigate how CME complexity varies with distance, for various propagation scenarios.
Since our simulations employ a spheromak magnetic structure for which rotations $\ge 180^\circ$ are expected for a large variety of spacecraft crossings (as shown in Appendix~\ref{sec:appendix_toy_model}), we have adapted the original classification to better distinguish rotations up to $360^\circ$.
We also further assign a numerical index ($\mathbb{C}$) to each ME class in order to rank the level of complexity of the detected structure.\ If an ME is detected at a spacecraft located at coordinates $(r, \theta, \phi)$, the following classification scheme is applied:
\begin{itemize}
\item $F_{270}$:\ ME signature with at least one component (i.e.\ $B_R$, $B_T$, or $B_N$) rotating $\ge 270^\circ$; complexity index $\mathbb{C}(r, \theta, \phi)=0$, corresponding to the least complex state.
\item $F_{180}$:\ ME signature with at least one component rotating $\ge 180^\circ$ and $< 270^\circ$; 
complexity index $\mathbb{C}(r, \theta, \phi)=1$.
\item $F_{90}$:\ ME signature with at least one component rotating $\ge 90^\circ$ and $< 180^\circ$; 
complexity index $\mathbb{C}(r, \theta, \phi)=2$.
\item $F_{30}$:\ ME signature with at least one component rotating $\ge 30^\circ$ and $< 90^\circ$; 
complexity index $\mathbb{C}(r, \theta, \phi)=3$.
\item $E$:\ ME signature with no component rotating $\ge 30^\circ$; 
complexity index $\mathbb{C}(r, \theta, \phi)=4$, corresponding to the most complex state.
\end{itemize}
Starting from the ME nominal boundaries, the ME start and end times are varied by $\pm 25\%$ of the total ME duration, in order to assess the variability of the classification with respect to slight variations of the boundaries \citep[reflecting the uncertainties in the boundary identification;][]{Riley2004, Al-Haddad2013}.\ The final ME classification is chosen as the most probable classification obtained among all possible combinations of boundaries.\ An example time series for a CME classified as having an $F_{270}$ ME signature, and the corresponding magnetic hodograms, are shown in Figure~\ref{fig:F270}. Additional examples are provided in Appendix~\ref{sec:appendix_example_time_series}.
\begin{figure*}
\centering
{\includegraphics[width=0.7\hsize]{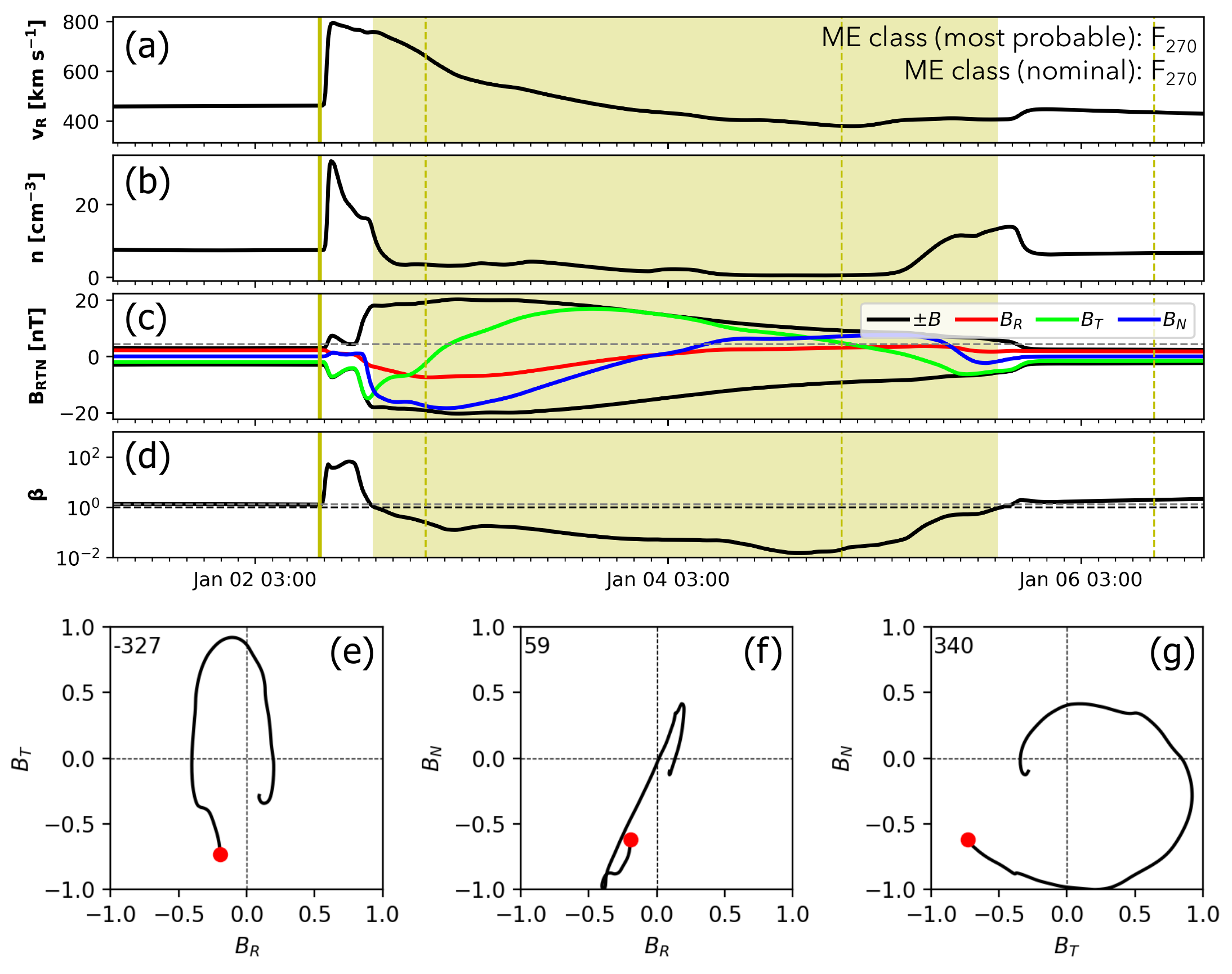}}
\caption{
$F_{270}$ ME type from run~A at $(r, \theta, \phi) = (1$~au, $0^\circ$, $-20^\circ$).
(a), (b), (c), (d): time series for the radial speed, number density, magnetic field components, and plasma $\beta$.
The continuous yellow line and dashed yellow area mark the shock-like disturbance and ME nominal boundaries, respectively.
Boundary variation ranges are marked by the dashed yellow lines.
(e), (f), (g): magnetic hodograms for nominal ME boundaries.  
The red dots mark the initial values of the magnetic field components.
} 
\label{fig:F270}
\end{figure*}

We include two additional categories representing the non-detection of an interplanetary shock-like perturbation (\textit{N}), 
and the detection of an interplanetary shock-like perturbation which was not followed by a ME (\textit{S}).\ In these cases, we do not assign a complexity index to the observed signatures because of their intrinsically different nature compared to ME signatures.\ 

{We tackle the known limitations of the spheromak model in reproducing CME global magnetic structures \citep[particularly with respect to stretched ``legs'' rooted to the Sun; e.g.][]{Scolini2019} by focusing the investigation of CME magnetic complexity to central regions only (Sections~\ref{subsec:results_classification} and \ref{subsec:results_changes}).
Furthermore, we note that the primary aim of this exploratory work is that of uncovering the complexity trends affecting CME structures during propagation through different solar wind structures, and that conclusive evidence for the applicability of our results to real events and possibly other flux rope configurations will have to be provided in future studies.}

\section{Results}
\label{sec:results}

\begin{figure*}
\centering
{\includegraphics[width=0.8\hsize]{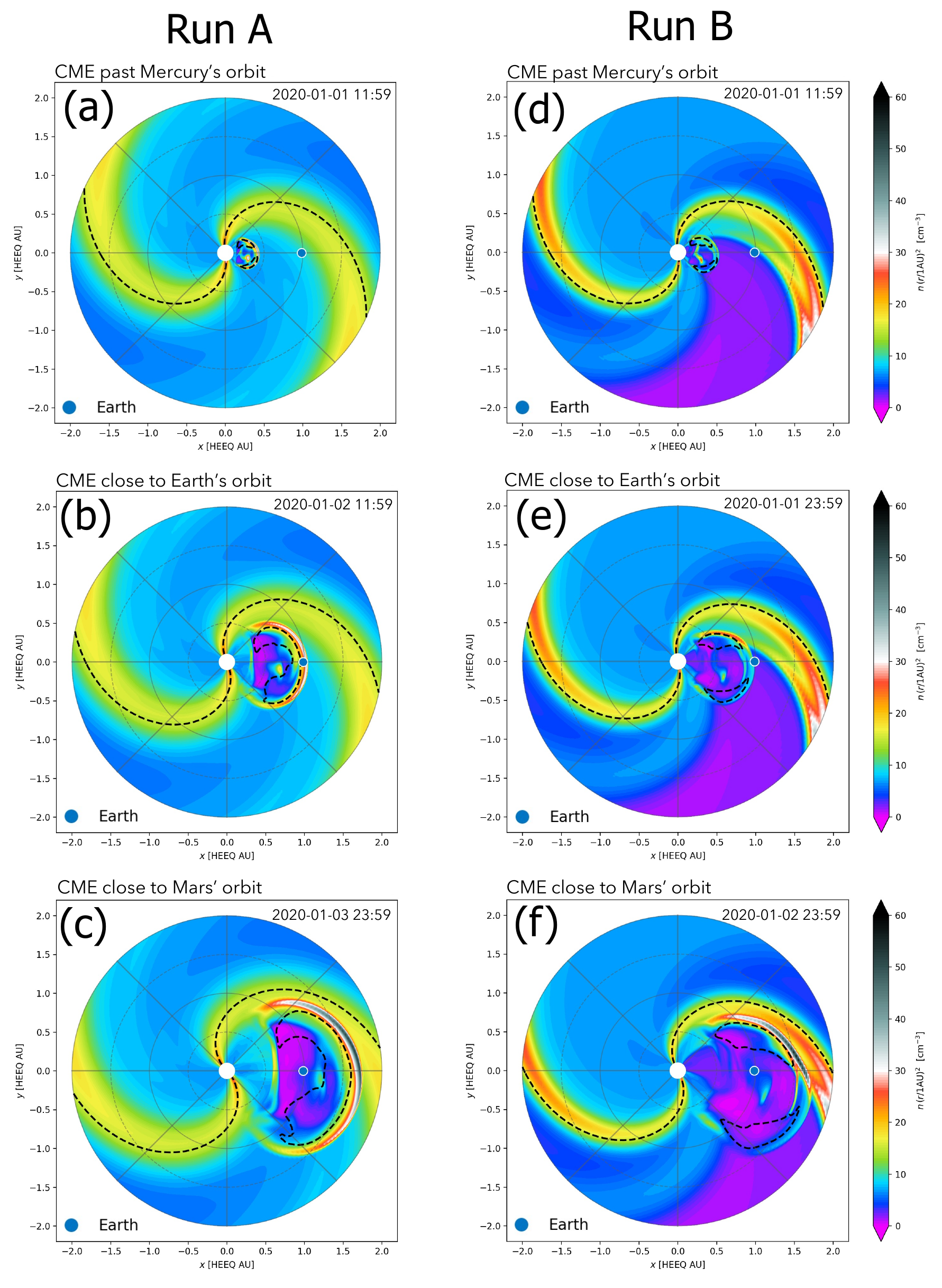}}
\caption{
EUHFORIA-simulated scaled number density ($n (r/1 \,\, \mathrm{au})^2$) in runs~A (left) and B (right), in the heliocentric equatorial plane.\ The different time steps show the CME fronts past Mercury's orbit ((a), (d)), close to Earth' orbit ((b), (e)), and close to Mars' orbit ((c), (f)).
The black dashed lines mark $B_R=0$~nT contours, including the position of the HCS.\
} 
\label{fig:equatorial}
\end{figure*}

Figure~\ref{fig:equatorial} provides an overview of the solar wind conditions and CME propagation in the heliocentric equatorial plane.\
%
As visible in the left column (run~A), if surrounded by a quiet and relatively homogeneous solar wind configuration, the CME structure propagates radially outwards without exhibiting any significant deflection or deformation of its front.\ This remains true even after the CME western portion enters a region of slightly slower speed and higher density associated with the HPS.\ Some deformation of the CME front is visible outside of the ecliptic plane associated with the crossing of the HPS, but as discussed in Sections~\ref{subsec:results_classification} and \ref{subsec:results_changes}, the effect on the detected magnetic complexity remains limited.\

The evolution of the CME large-scale structure appears significantly different when interacting with a HSS and SIR (Figure~\ref{fig:equatorial}, right column).\ In run~B, the CME is inserted in the heliospheric domain across the HSS.\ In the early phase of interplanetary propagation, the CME propagates through the HSS, while the HCS/HPS is located west of it.\ The western flank of the CME starts interacting with the SIR within 0.5~au, and the interaction progressively encompasses larger portions of the CME becoming dominant beyond 1~au.\ The large pressure associated with the SIR blocks the westernmost part of the CME front, deflecting the CME towards the east as visible in panels (e) and (f) in Figure~\ref{fig:equatorial}, consistent with expectations \citep[e.g.][]{Wang2004}.\
Recent observational studies suggest this configuration should lead to the development of a higher CME magnetic complexity than run~A, particularly due to the CME interaction with the SIR \citep[as shown by][]{Winslow2021}.\ 
It is also important to mention that in run~B, the CME is launched north of the HCS/HPS (Figure~\ref{fig:solar_wind}(d)), while in run~A, the CME was inserted right across it (Figure~\ref{fig:solar_wind}(b)).\ 
The HCS/HPS is therefore expected to affect the CME structure more in run~A than in run~B \citep[][]{Winslow2016}.\
The extent of the alterations induced by these solar wind structures on the CME are investigated more in-depth through statistical methods, and are discussed in Sections~\ref{subsec:results_classification} and \ref{subsec:results_changes}.\

\subsection{Spatial distribution of CME complexity as a function of distance}
\label{subsec:results_classification}

\begin{figure*}
\centering
{\includegraphics[width=0.8\hsize]{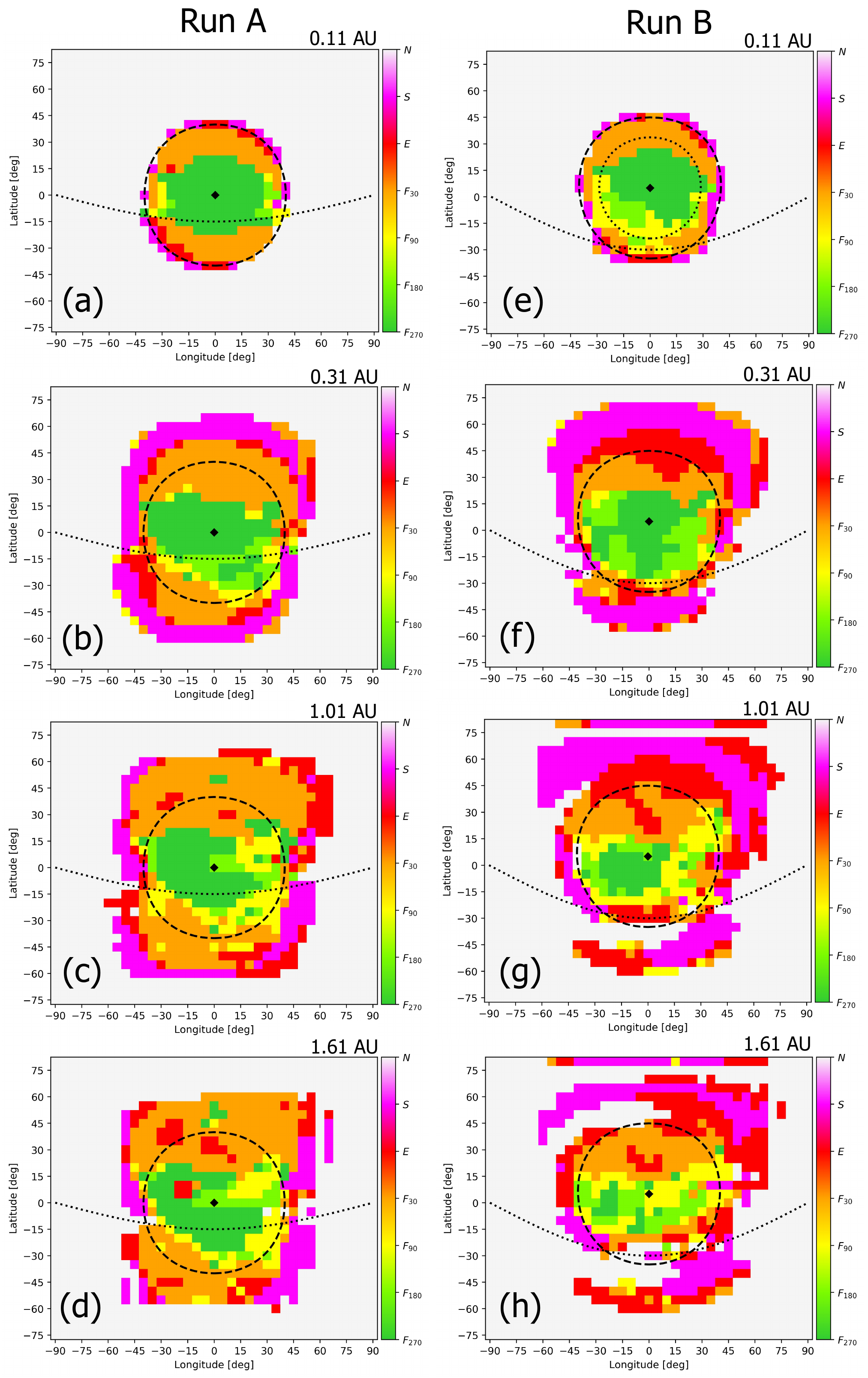}}
\caption{
Distribution of CME signatures
from runs~A (left) and B (right), at  0.1~au ((a), (e)), 0.3~au ((b), (f)), 1~au ((c), (g)), and 1.6~au ((d), (h)).
Diamond: initial CME direction.\ Dashed black line: initial CME cross-section.\ Dotted black lines: HCS and HSS initial locations.
} 
\label{fig:classification}
\end{figure*}

The results of the identification and classification analysis introduced in Section~\ref{subsec:methods_classification} are provided in Figure~\ref{fig:classification}, shown as longitude--latitude maps colored by classification type, for selected heliocentric distances.

As expected, immediately after insertion at 0.1~au, the extent of the CME front in the two simulations is very similar, and it matches well the nominal CME cross-section expected from the initial half angular width of $45^\circ$ (Figure~\ref{fig:classification}, top row).\ At this very early stage, the spatial distribution of the shock-like and ME classifications is also comparable between runs~A and B, and is qualitatively consistent with the results expected for a spheromak structure (Figure~\ref{fig:spheromak_toy_model}).\ 

At 0.3~au (Figure~\ref{fig:classification}, second row) the CME cross-section has expanded to an effective half-width of $\sim 60^\circ$, indicating an over-expansion in the early stage of the propagation \citep[][]{Scolini2019, Scolini2020A&A}.
By the time the CME reaches 0.3~au, significant differences are visible in the CME cross-section and in the spatial distribution of the classification types between the two runs.\ While run~A preserves a relatively-symmetric classification distribution with respect to the spheromak main axis (lying parallel to the ecliptic plane and the local HCS direction), dominated by $F_{30}$ and $F_{270}$ types, run~B exhibits a distorted cross-section and increased $E$ ejecta types.
At this stage, an increased detection of shock-only ($S$) signatures close to the CME flanks indicates the formation of a CME-driven shock and sheath region that is more extended than its driver \citep{Kilpua2017}.

In both runs, the CME cross-section at 1~au and 1.6~au (Figure~\ref{fig:classification}, third and fourth rows) remains similar to that at 0.3~au, meaning the angular expansion of the CME is almost negligible beyond Mercury's orbit.\
On the contrary, beyond 0.3~au the spatial distribution of the ME classification types becomes visibly less regular with heliocentric distance, especially for run~B.
The differences between runs~A and B at this late stage of propagation are remarkable: in run~A, the CME cross-section remains quasi-circular, despite a shrinking in the detection of shock-like signatures around its flanks.\ The classification in the CME core region also remains largely unchanged (Figure~\ref{fig:classification}(c) and (d)).
In run~B (Figure~\ref{fig:classification}(g) and (h)), a similar shrinking in the detection of shock-like signatures is visible around the CME flanks, particularly in the regions affected by the interaction with the HSS (north-east) and HCS/HPS (south-east).\ However, a decrease of the less complex ME types ($F_{270}$ and $F_{180}$ classes) in favor of more complex ones ($F_{90}$, $F_{30}$, and $E$ classes) is observed in the core CME region.\ The irregular spatial distribution of ME types in this region is also associated with a high probability for an inner and outer spacecraft along a given ($\theta, \phi$) direction to detect different ME types, as further discussed in Section~\ref{subsec:results_changes}.
We note that this is partly due to the formation of a CME--SIR merged interaction region \citep{Rouillard2010}, which affects the efficiency of the ME detection algorithm at larger distances.\ Similar difficulties are likely to affect the identification of MEs from actual in situ data, thereby making this limitation particularly instructive also with respect to future observational applications.\

The different CME evolutionary behaviors identified above result in different probabilities to detect the various ME classes and additional $S$ and $N$ signatures as a function of the heliocentric distance in the two simulations performed.\ 
{Considering only spacecraft crossings within $45^\circ$ from the CME initial direction (i.e.\ well within the CME effective half angular width of $\sim 60^\circ$ reached at 0.3~au, corresponding to relatively central impact locations), we find that r}un~A is dominated by the detection of $F_{270}$ and $F_{30}$ ejecta types, and more than $90$\% of all spacecraft detect the passage of an $F$ type.\ 
Notably, these probabilities are mostly independent from heliocentric distance, and remain consistent with those expected for a spheromak structure (i.e.\ not interacting with the solar wind, as shown in Figure~\ref{fig:spheromak_toy_model}).\ 
In run~B, $F_{30}$ detections dominate all heliocentric distances, while the second most-detected ejecta type passes from $F_{270}$ to $F_{90}$ beyond 0.8~au.\ A more-than-doubled fraction of $E$, $S$ and $N$ complex types ($> 20$\%) is also observed. 
The probabilities for run~B also exhibit a strong dependence on the heliocentric distance:\
all signature types are almost equally represented by the time the CME reaches 1.6~au,
as opposed to run~A where a clearly bi-modal distribution is preserved during propagation.\

\subsection{Probability of detecting CME complexity changes with distance}
\label{subsec:results_changes}

\begin{figure*}
\centering
{\includegraphics[width=0.9\hsize]{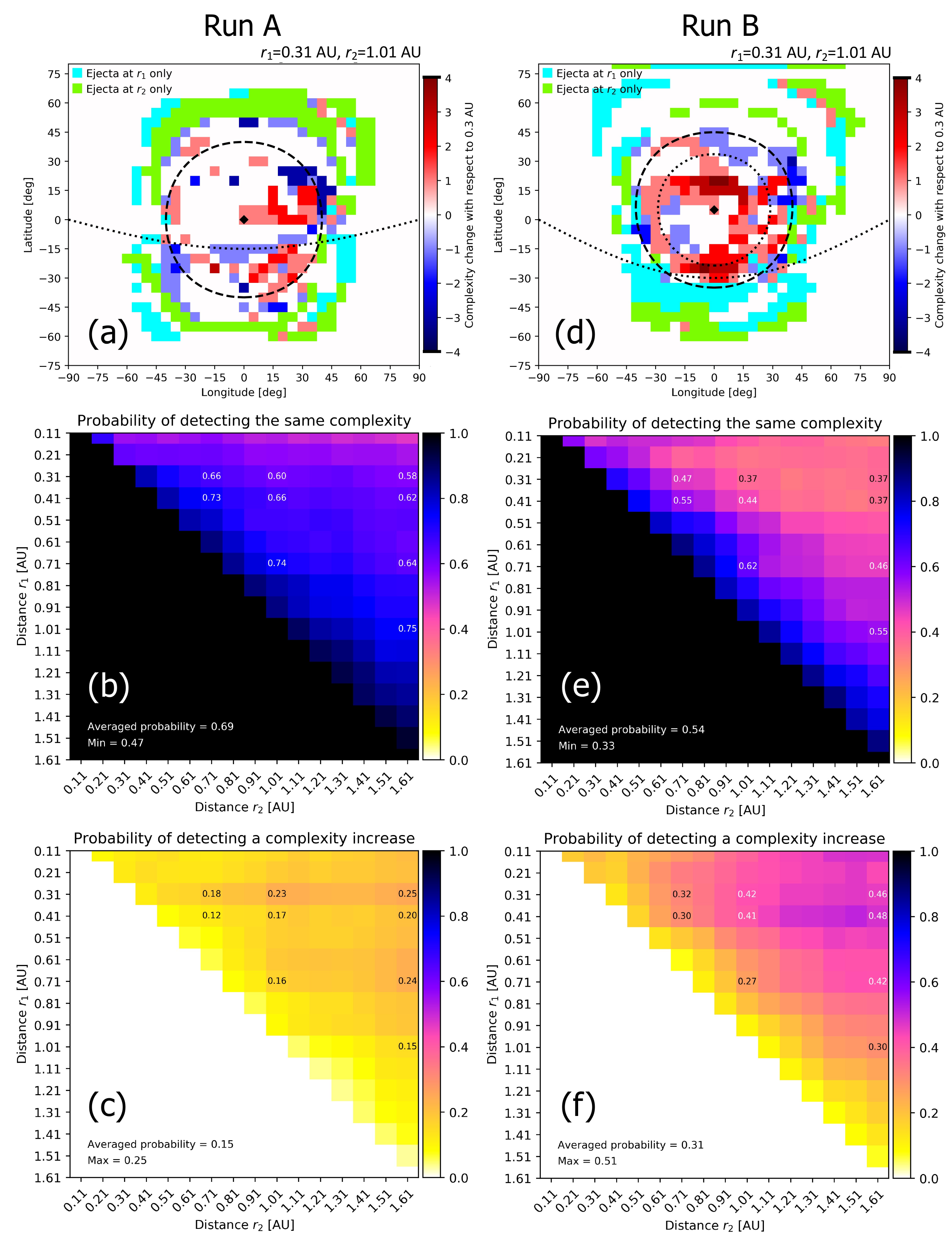}}
\caption{
(a), (d): Distribution of magnetic complexity changes between $0.3$~au and $1$~au in runs~A (left) and B (right). The color scale quantifies $\Delta \mathbb{C}(r_1, r_2, \theta, \phi)$ along directions detecting an ME at both distances. Green (cyan) squares indicate directions where only the inner (outer) spacecraft detected an ME.
(b), (c), (e), (f): Probability of detecting an unchanged or increased CME complexity at spacecraft in radial alignment, for different distance pairs.\ 
Only spacecraft pairs within $45^\circ$ from the initial CME direction, and that detected an ME at both distances, are considered.
The ``averaged probability'' is the sum of all matrix elements normalized by the number of elements, i.e.\ 120.
} 
\label{fig:change}
\end{figure*}

To quantify the overall change in CME complexity with heliocentric distance, we first consider a generic pair of heliocentric distances, i.e.\ $r_1$ and $r_2$, with $r_1 < r_2$. Then, we consider all pairs of virtual spacecraft in radial alignment located within $45^\circ$ from the CME initial direction (justified by the CME initial half-width {and the need to restrict ourselves to central CME regions, so to limit the impact of spheromak limitations around the flanks}), and that detected an ME signature at both distances.\ 
Moving along fixed $(\theta^*, \phi^*)$ directions satisfying the above criteria, 
we compute the changes in CME complexity between the inner spacecraft at $r_1$ and the outer spacecraft at $r_2$, as 
$\Delta \mathbb{C}(r_1, r_2, \theta^*, \phi^*) = \mathbb{C}(r_2,  \theta^*, \phi^*) - \mathbb{C}(r_1, \theta^*, \phi^*)$.
$\Delta \mathbb{C}(r_1, r_2, \theta^*, \phi^*) >, =,$ or $< 0$ indicates directions where an increased, unchanged, or decreased CME complexity with heliocentric distance was detected.
The results of this procedure are provided in Figure~\ref{fig:change} (top row) for the notable case of radially aligned spacecraft at 0.3~au (consistent with the orbit of Mercury and Solar Orbiter's perihelion), and at 1~au (consistent with Earth's orbit).

By counting how many directions detected a increased, unchanged, or decreased complexity, we determine the overall probability (normalized between $0$ and $1$) to detect complexity changes between two spacecraft in radial alignment at distances $r_1$ and $r_2$.\ Finally, by applying the same process to all distance pairs $(r_1, r_2)$, we construct global matrix plots, as shown in Figure~\ref{fig:change} (second and third rows).\ 

The top row in Figure~\ref{fig:change} exemplifies how the CME in run~B develops a higher complexity as it propagates, compared to run~A.
In run~A (panel (a)), a predominance of $\Delta \mathbb{C} = 0$ white squares indicating a stable complexity is visible within $45^\circ$ from the initial CME direction.\ A quasi-symmetric distribution of the classification changes beyond $45^\circ$ is also visible, where spacecraft pairs detecting an ME at the outer spacecraft only (in green) reflect the slight ME expansion between 0.3~au and 1~au.\ Cyan squares are likely the results of spurious ME detection at 0.3~au.
Run~B (panel (b)), on the other hand, shows a predominance of $\Delta \mathbb{C} >0$ red squares within the nominal ($45^\circ$) initial CME cross-section.\ 
The presence of green squares within a $60^\circ$ range from the CME center indicates a slight CME expansion in the northern hemisphere compared to 0.3~au, while the presence of cyan squares immediately south of the CME and HCS/HPS suggests some erosion of the ME flanks during propagation from 0.3 to 1~au.\ 

Overall, run~A supports the idea that CMEs propagating through a quiet heliosphere tend to maintain their complexity relatively unchanged.\ The most likely scenario is that the ME classification made at an inner spacecraft remains the same at a radially aligned outer spacecraft (Figure~\ref{fig:change}(b)), regardless of their radial separation (average probability of $69$\%, minimum probability of $48$\%).\ 
All the most notable alignment configurations (i.e.\ involving Mercury, Venus, spacecraft at 1~au, and Mars) have a probability between $58$\% (for a Mercury--Mars conjunction) and $75$\% (for a 1~au--Mars conjunction) to detect exactly the same ME type at both distances. Unsurprisingly, the larger the radial separation between the spacecraft, the lower this probability is.\ Furthermore, the larger the heliocentric distance of the inner spacecraft, the higher the probability is to detect the same complexity level at outer distances, indicating that changes are more likely to occur closer to the Sun.
The opposite trend is observed in Figure~\ref{fig:change}(c), showing that although the probability to detect a complexity increase for run~A becomes larger with larger radial separations, it remains quite modest (average probability of $15$\%, maximum probability of $25$\%). Complexity decreases (not shown) are similarly unlikely (average probability of $16$\%, maximum probability of $34$\%).

Conversely, in run~B the indicators used to evaluate CME complexity changes show that, on average, CMEs propagating through a structured solar wind still tend to preserve their complexity (average probability of $54$\%; Figure~\ref{fig:change}(e)), but their probability to transition to a more complex configuration is more than doubled compared to run~A (average probability of $31$\%; Figure~\ref{fig:change}(f)).\ Results vary with distance, as indicated by a minimum (maximum) probability of $33$\% ($51$\%) to detect an unchanged (increased) magnetic complexity.
Furthermore, the probability to detect the same ME type at two different distances is $>50$\% for only three of the notable radial alignments considered: Mercury--Venus, Venus--1~au, and 1~au--Mars. In all other cases, the most likely scenario is that two locations detect different ME types. Specifically, the probability to detect a complexity increase ranges from $27$\% for a Venus--1~au conjunction, to $48$\% for a Mercury--Mars conjunction.
Furthermore, in the case of run B, complexity increases are more prominent when considering the alignment of a spacecraft located within 0.4~au from the Sun, with one beyond 1~au, reflecting the interaction of the CME with the preceding SIR over a longer distance range, as shown in Figure~\ref{fig:equatorial}(e), (f).
Finally, the probability to detect complexity decreases in run~B remains similar to run~A.

Based on our numerical investigation, we conclude that the interaction with solar wind structures, and particularly SIRs, can double the probability for a CME to increase its magnetic complexity.\ Most importantly (as illustrated in Figures~\ref{fig:classification} and \ref{fig:change}), this result does not depend on the distance of the spacecraft crossing from the ME center. As such, changes of magnetic complexity detected by spacecraft that are in exact radial alignment are likely signs of interaction with other structures, rather than inherent to the CME evolution even if crossed far from the center.\

\section{Discussion and conclusions}
\label{sec:conclusions}

We performed a numerical study of CMEs interacting with different solar wind configurations, with the scope of determining under which conditions and to what extent CMEs exhibit variations of their magnetic structure and complexity during propagation through interplanetary space.\ We {employed a novel modeling approach to assess} the probability to detect changes in CME magnetic complexity by using a swarm of {simulated} spacecraft in radial alignment given the absence/presence of corotating structures. {We restricted our attention to the central part of the CME structure in order to limit the effect of known limitations arising from the use of a spheromak CME model.}
 
From the comparative analysis of non-interacting/interacting scenarios, distinct evolutionary behaviors characterizing CMEs propagating through different ambient conditions have emerged.
Our results provide evidence that the interaction with such structures, and particularly with SIRs, can double the probability for a CME to increase its magnetic complexity during propagation.\ This result is independent from the CME impact angle, suggesting that the detection of complexity changes is likely an indication of interactions with other structures, rather than the result of a crossing far from the CME center.
{The present work provided first evidence that CME structures propagating through different solar wind backgrounds develop different complexity evolutionary patterns, based on numerical simulations employing a spheromak flux rope model. Conclusive evidence that similar trends apply to real events and different flux rope models should be assessed in future studies.}

{Another way to look at the development of CME magnetic complexity during propagation involves consideration of the coherence of the magnetic structure as a function of heliocentric distance \citep{Owens2017, Lugaz2018}. In our simulations, we find that the Alfv\'{e}n speed in the ME at 1~au is $\sim 120$~km~s$^{-1}$ ($\sim 97$~km~s$^{-1}$) in run~A (B). While a detailed investigation of CME coherence goes beyond the scope of this work, we note that these values are comparable to those reported in Figure~2(c) by \citet{Owens2017}, and that the higher Alfv\'{e}n speed retrieved in run~A may indicate a more coherent evolution of the ME structure than in run~B, in agreement with our results of the complexity changes highlighted in Figures~\ref{fig:classification} and \ref{fig:change}}.
{Future works extending beyond this first exploratory investigation are needed in order to draw more general conclusions on this topic.}

This work represents the first attempt to quantify complexity changes in CME magnetic structures using numerical simulations.\ 
Our simulations assumed very idealized solar wind conditions and did not include any latitudinal dependence of its properties, enabling us to quantify the effect of the presence of a HSS and SIR on CME complexity, and the comparability of different runs.\ The results presented here shall {therefore} be interpreted as lower limits, as ubiquitous distortions of the local solar wind properties \citep[e.g.][]{Torok2018} {are} likely {to} induce higher complexity changes in real CMEs {both in the absence and presence} of SIRs along their path.\ {Investigations exploring a broader range of CME--solar wind interactions,} the spatial dependence of CME complexity and its changes, and the comparison with observations for real events{, will be explored in future studies}.\

Our results shed new light on the evolution of CME magnetic structures, helping the identification and interpretation of CME conjunction observations involving both past and current missions (such as Parker Solar Probe, Solar Orbiter, and BepiColombo), and providing guidelines for the planning of future missions involving alternative alignment configurations.\ 


\acknowledgments 
C.S. acknowledges the NASA Living With a Star Jack Eddy Postdoctoral Fellowship Program, administered by UCAR's Cooperative Programs for the Advancement of Earth System Science (CPAESS) under award no.\ NNX16AK22G.
R.M.W. acknowledges support from NASA grant 80NSSC19K0914. 
N.L. acknowledges support from NASA grant 80NSSC20K0700. 
S.P. has received funding from the European Union's Horizon 2020 research and innovation program under grant agreement No 870405 (EUHFORIA 2.0), C14/19/089 (KU Leuven), G.0D07.19N (FWO-Vlaanderen), SIDC Data Exploitation (ESA Prodex-12), and the Belspo projects BR/165/A2/CCSOM and B2/191/P1/SWiM.
EUHFORIA is developed as a joint effort between the University of Helsinki and KU Leuven.
The simulations were carried out at the VSC -- Flemish Supercomputer Center, funded by the Hercules foundation and the Flemish Government -- Department EWI.

%






\appendix

\section{Spheromak magnetic structure}
\label{sec:appendix_toy_model}

\begin{figure*}
\centering
{\includegraphics[width=0.9\hsize]{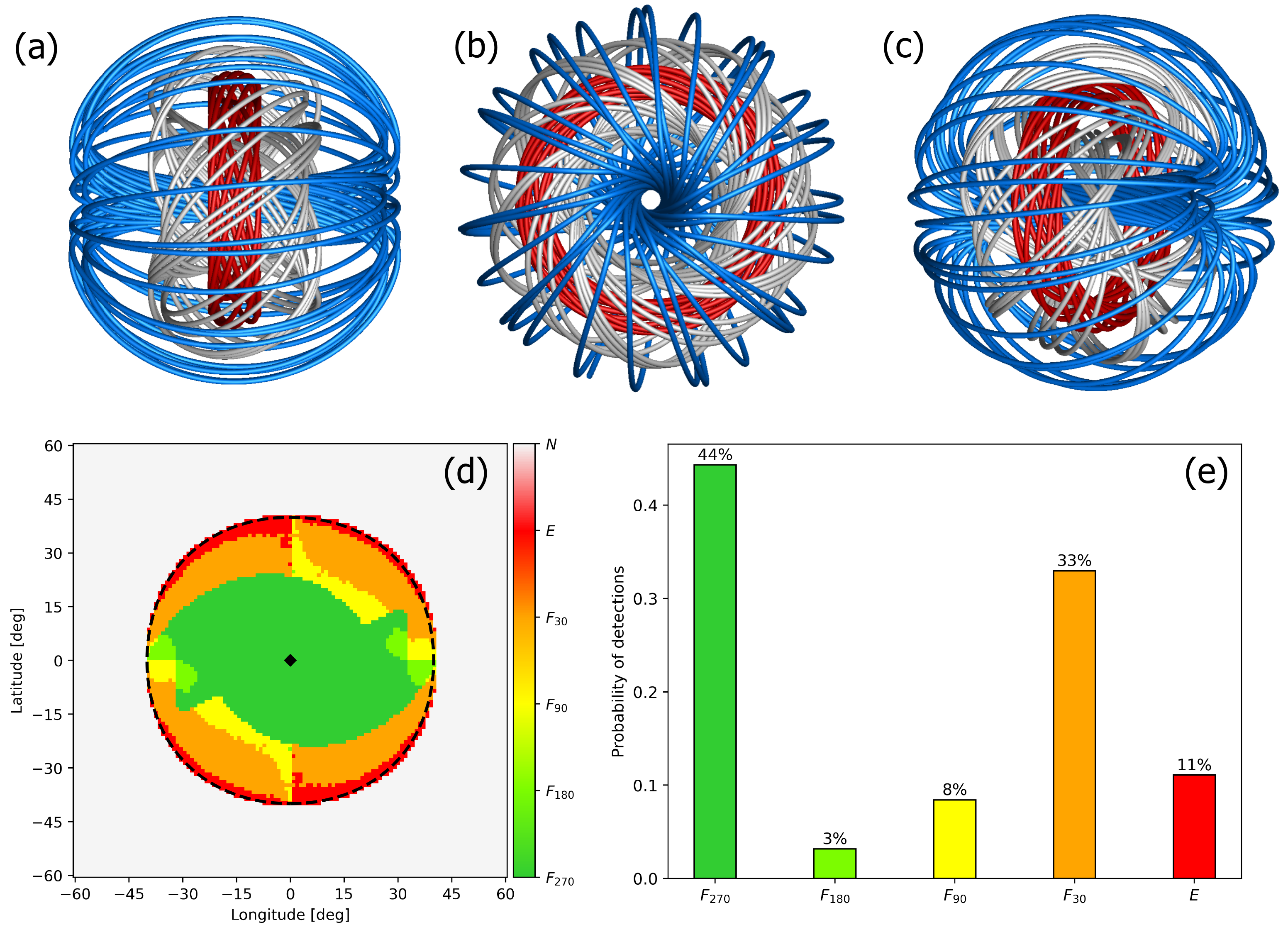}}
\caption{
Spheromak magnetic structure used in this work's simulations.
Top: 3D visualization of selected magnetic fields from three perspectives: front view (a), side view (b), angled view (c).
Different colors mark field lines characterized by different morphologies.
(d): Longitude--latitude 2D map showing the spatial distribution of the ME signature classifications introduced in Section~\ref{subsec:methods_classification}, obtained by radially crossing the structure at various impact angles.
(e): Probability to detect the different ME signatures considering all possible radial crossings throughout the structure.
} 
\label{fig:spheromak_toy_model}
\end{figure*}

Figure~\ref{fig:spheromak_toy_model} provides an overview of the nominal, stand-alone magnetic field structure of the linear force-free spheromak model used in runs~A and B, i.e.\ without accounting for any interaction with the solar wind. 
As visible from panels (a)--(c), all magnetic field lines are confined within a spherical surface. 
Panel~(a) provides a frontal view of the structure, similar to the one that would have been seen by an observer located near the ecliptic plane along the Sun--Earth line in run~A before the insertion of the CME into the heliospheric domain. 
Panels~(b) and (c) provide additional side and angled views of the structure.
By crossing the spheromak structure in the radial direction at various impact angles (varying the crossing directions by $\Delta \theta = \Delta \phi =1^\circ$ incremental steps) and calculating the rotations of the magnetic field components along each direction, we generate a longitude--latitude 2D spatial distribution map of the ejecta classifications introduced in Section~\ref{subsec:methods_classification} (Figure~\ref{fig:spheromak_toy_model}(d)).
The distribution is symmetric with respect to the spheromak main axis, oriented along the equatorial plane, and is dominated by a central core of $F_{270}$ ejecta types surrounded by $F_{30}$ classifications in the north and south flank regions. Minor contributions from $F_{180}$ and $F_{90}$ types are visible in the core-to-flank transition region, while $E$ types are detected only by crossings at the very edge of the structure.
Figure~\ref{fig:spheromak_toy_model}(e) summarizes the probability to the detect the different ME types over the totality of crossings considered. The detection probability is highest for $F_{180}$ ($44$\%) and $F_{30}$ ($33$\%) types, and significantly lower for the remaining types ($11$\%, $8$\%, $3$\% for $E$, $F_{90}$, $F_{180}$ types, respectively).

\section{Interplanetary CME identification algorithm}
\label{sec:appendix_algorithms}

\subsection{Shock-like signatures identification algorithm} 
\label{subsec:appendix_algorithms_shock}

At each virtual spacecraft in the simulation domain, we determine the arrival time of the shock-like CME-driven perturbation by scanning the radial speed, density, and magnetic field time series (having a cadence of $\Delta t = 10$~min) forward in time, and applying the following conditions:
\begin{equation}
\overline{v_R}(t_{i}:t_{i}+1 \, \mathrm{hr}) - \overline{v_R}(t_{i}-1 \, \mathrm{hr}:t_{i}) \ge 20 \,\, \mathrm{km \,\, s^{-1}} \,\,\,\, \land \,\,\,\, \overline{n}(t_{i}:t_{i}+1 \, \mathrm{hr}) \ge  \overline{n}(t_{i}-1 \, \mathrm{hr}:t_{i})
\,\,\,\, \land \,\,\,\, \overline{B}(t_{i}:t_{i}+1 \, \mathrm{hr}) \ge  \overline{B}(t_{i}-1 \, \mathrm{hr}:t_{i}),
\label{eqn:shock}
\end{equation}
where $t_i$ is a generic time in the time series, 
and $\overline{X}(t_{i}:t_{j})$ is the average of quantity $X$ calculated between time $t_i$ and time $t_j$.
The arrival time of the CME-driven perturbation, $t_{sh}$, is determined as the first time $t_i$ at which the system of equations~\ref{eqn:shock} is satisfied.
These conditions, which we have verified visually for selected spacecraft at different heliocentric distances from the Sun, are adapted versions of the speed, density, and magnetic field conditions used to detect fast-forward interplanetary shocks from in situ solar wind measurements at 1~au employed by the Database of Heliospheric Shock Waves \citep[\url{http://www.ipshocks.fi};][]{Kilpua2015} and by the ACE Real-Time Shock database \citep[\url{http://www.srl.caltech.edu/ACE/ASC/DATA/Shocks/shocks.html};][]{Vorotnikov2008}.

\subsection{Magnetic ejecta identification algorithm} 
\label{subsec:appendix_algorithms_ejecta}

At locations where a CME-driven perturbation is detected, time series are scanned in order to assess the presence of a ME after the shock-like discontinuity.\
To do so, we first define the average interplanetary magnetic field ($B_{sw}$) and solar wind plasma $\beta$ ($\beta_{sw}$) in the 6 hours prior to the arrival time of the shock-like CME driven perturbation, as:
\begin{equation}
    B_{sw} = \overline{B}(t_{sh}-6 \, \mathrm{hr}:t_{sh}), \,\,\,\,
    \beta_{sw} = \overline{\beta}(t_{sh}-6 \, \mathrm{hr}:t_{sh}).
\label{eqn:sw_conditions}
\end{equation}
Depending on the $\beta_{sw}$ recovered at a given spacecraft position (which depends on its heliocentric distance and on the local solar wind conditions), we then consider separately the cases of a magnetically-dominated ($\beta_{sw} \le 1$) and a plasma-dominated ($\beta_{sw} > 1$) environment.

\medskip
\paragraph{Low-$\beta$ solar wind}

For a magnetically-dominated $\beta_{sw} \le 1$ solar wind, we scan the magnetic field and plasma $\beta$ time series forward in time starting from $t_{sh}$, and apply the following conditions:
\begin{equation}
    \overline{B}(t_{i}:t_{i}+1 \, \mathrm{hr}) \ge 1.5 \, B_{sw} \,\,\,\, \land \,\,\,\,
    \overline{\beta}(t_{i}:t_{i}+1 \, \mathrm{hr}) \le \beta_{sw}.
\label{eqn:me_start_low_beta}
\end{equation}
The start time of the ME, $t_{start}$, is determined as the first time $t_i$ at which Equation~\ref{eqn:me_start_low_beta} is satisfied.\ 
%
Only if an ME start is detected at a given location, do 
we continue with the determination of the ME end time.\ 
To do so, we scan the magnetic field time series forward in time starting from $t_{start}$, and apply the following conditions:
\begin{equation}
    \overline{B}(t_{i}-1 \, \mathrm{hr}:t_{i}) \ge 1.5 \, B_{sw} \,\,\,\, \land \,\,\,\,
    \overline{B}(t_{i}:t_{i}+1 \, \mathrm{hr}) \le 1.5 \, B_{sw},
\label{eqn:me_end_low_beta}
\end{equation}
to detect when the magnetic field drops below $1.5 \, B_{sw}$, which we have taken as threshold condition to characterize the boundary of the ME in both Equations~\ref{eqn:me_start_low_beta} and \ref{eqn:me_end_low_beta}.\ 
This threshold value was chosen after having tested values between $1.2 \, B_{sw}$ and $1.6 \, B_{sw}$, and having verified visually that it provided the best compromise, i.e.\ minimizing the number of false positive/negative ME detections.
The end time of the ME, $t_{end}$, is determined as the first time $t_i$ at which Equation~\ref{eqn:me_end_low_beta} is satisfied.
We verified visually that Equations~\ref{eqn:me_start_low_beta} and~\ref{eqn:me_end_low_beta} gave reasonable and consistent results throughout the whole range of heliocentric distances sampled by the virtual spacecraft in the model domain.\ A low $\beta$ condition to determine the end of the ME was also tested, but was found to perform less reliably than Equation~\ref{eqn:me_end_low_beta}, which is uniquely based on the magnetic field strength.\

\medskip
\paragraph{High-$\beta$ solar wind}

For a plasma-dominated $\beta_{sw} > 1$ solar wind, we scan the magnetic field and plasma $\beta$ time series forward in time starting from $t_{sh}$, and impose high magnetic field and low $\beta$ conditions to identify the start of the ME: 
\begin{equation}
    \overline{B}(t_{i}:t_{i}+1 \, \mathrm{hr}) \ge 1.5 \, B_{sw} \,\,\,\, \land \,\,\,\,
    \overline{\beta}(t_{i}:t_{i}+1 \, \mathrm{hr}) \le 1.
\label{eqn:me_start_high_beta_1}
\end{equation}
The start time of the ME, $t_{start}$, is determined as the first time $t_i$ at which Equation~\ref{eqn:me_start_high_beta_1} is satisfied.\ 
%
Only if an ME start is detected at a given location via Equations~\ref{eqn:me_start_high_beta_1}, do
we continue with the determination of the ME end time.
As the determination of the end time proved to be a more complex task than the identification of the start time, two alternative conditions based on the magnetic field and plasma $\beta$ are applied, to account for the variety of plasma properties encountered.
In particular, we scan the magnetic field time series forward in time starting from $t_{start}$, and apply the following conditions:
\begin{equation}
    \overline{B}(t_{i}:t_{i}+1 \, \mathrm{hr}) < 1.5 \, B_{sw} \,\,\,\, \land \,\,\,\,
    \overline{\beta}(t_{i}-1 \, \mathrm{hr}:t_{i}) \le 1,
\label{eqn:me_end_high_beta_1a}
\end{equation}
or, alternatively,
\begin{equation}
    \overline{B}(t_{i}-1 \, \mathrm{hr}:t_{i}) \ge 1.5 \, B_{sw} \,\,\,\, \land \,\,\,\,
    \overline{\beta}(t_{i}:t_{i}+1 \, \mathrm{hr}) > 1.
\label{eqn:me_end_high_beta_1b}
\end{equation}
Equation~\ref{eqn:me_end_high_beta_1a} identifies the end boundary of the ME based on a low magnetic field condition, while Equation~\ref{eqn:me_end_high_beta_1b} is based on a high $\beta$ condition.\ The end time of the ME, $t_{end}$, is determined as the first time $t_i$ at which Equations~\ref{eqn:me_end_high_beta_1a} or \ref{eqn:me_end_high_beta_1b} are satisfied.
We verified visually that Equations~\ref{eqn:me_start_high_beta_1}--\ref{eqn:me_end_high_beta_1b} gave reasonable and consistent results throughout the whole range of heliocentric distances sampled by the virtual spacecraft in the model domain.

A further visual inspection of the results assessed there were cases where the ejecta could be recognized to cross a virtual spacecraft by eye, but which the conditions in Equations~\ref{eqn:me_start_high_beta_1}--\ref{eqn:me_end_high_beta_1b} failed to identify due to the CME plasma $\beta$ being lower than $\beta_{sw}$, but higher than 1.\ To account for these additional cases, a secondary identification of the ME start time is performed by applying the following criteria:
\begin{equation}
    \overline{B}(t_{i}:t_{i}+1 \, \mathrm{hr}) \ge 1.5 \, B_{sw} \,\,\,\, \land \,\,\,\,
    \overline{\beta}(t_{i}:t_{i}+1 \, \mathrm{hr}) \le \beta_{sw}.
\label{eqn:me_start_high_beta_2}
\end{equation}
%
Only if an ME start is detected at a given location via Equations~\ref{eqn:me_start_high_beta_2}, do
we continue with the determination of the ME end time.
Similarly to the case above, two alternative conditions based on the magnetic field and plasma $\beta$, are applied to detect the trailing edge of the ME:
\begin{equation}
    \overline{B}(t_{i}:t_{i}+1 \, \mathrm{hr}) < 1.5 \, B_{sw} \,\,\,\, \land \,\,\,\,
    \overline{\beta}(t_{i}-1 \, \mathrm{hr}:t_{i}) \le \beta_{sw},
\label{eqn:me_end_high_beta_2a}
\end{equation}
or, alternatively,
\begin{equation}
    \overline{B}(t_{i}-1 \, \mathrm{hr}:t_{i}) \ge 1.5 \, B_{sw} \,\,\,\, \land \,\,\,\,
    \overline{\beta}(t_{i}:t_{i}+1 \, \mathrm{hr}) > \beta_{sw}.
\label{eqn:me_end_high_beta_2b}
\end{equation}
Also in this case we visually inspected the classification resulting from the application of Equations~\ref{eqn:me_start_high_beta_2}--\ref{eqn:me_end_high_beta_2b} verifying their reliability and consistency throughout the whole range of heliocentric distances sampled by the virtual spacecraft in the model domain.


\section{Example time series}
\label{sec:appendix_example_time_series}

\begin{figure*}
\centering
{\includegraphics[width=0.7\hsize]{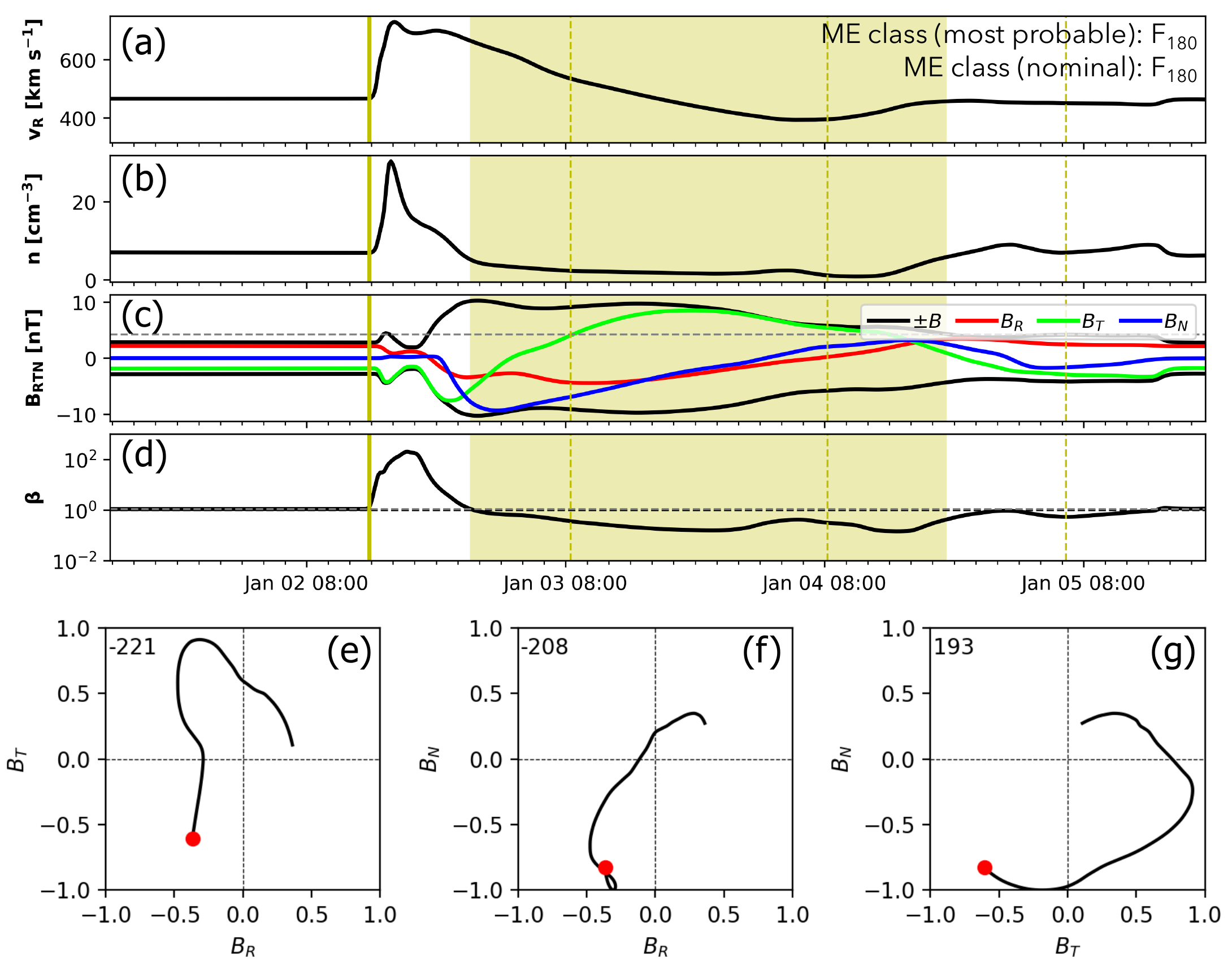}}
\caption{
Example of ME identification and classification from run~A at $r=1$~au, $\theta=5^\circ$, $\phi=-35^\circ$ ($F_{180}$ class).
The panels show the same quantities as in Figure~\ref{fig:F270}.
} 
\label{fig:F180}
\end{figure*}

\begin{figure*}
\centering
{\includegraphics[width=0.7\hsize]{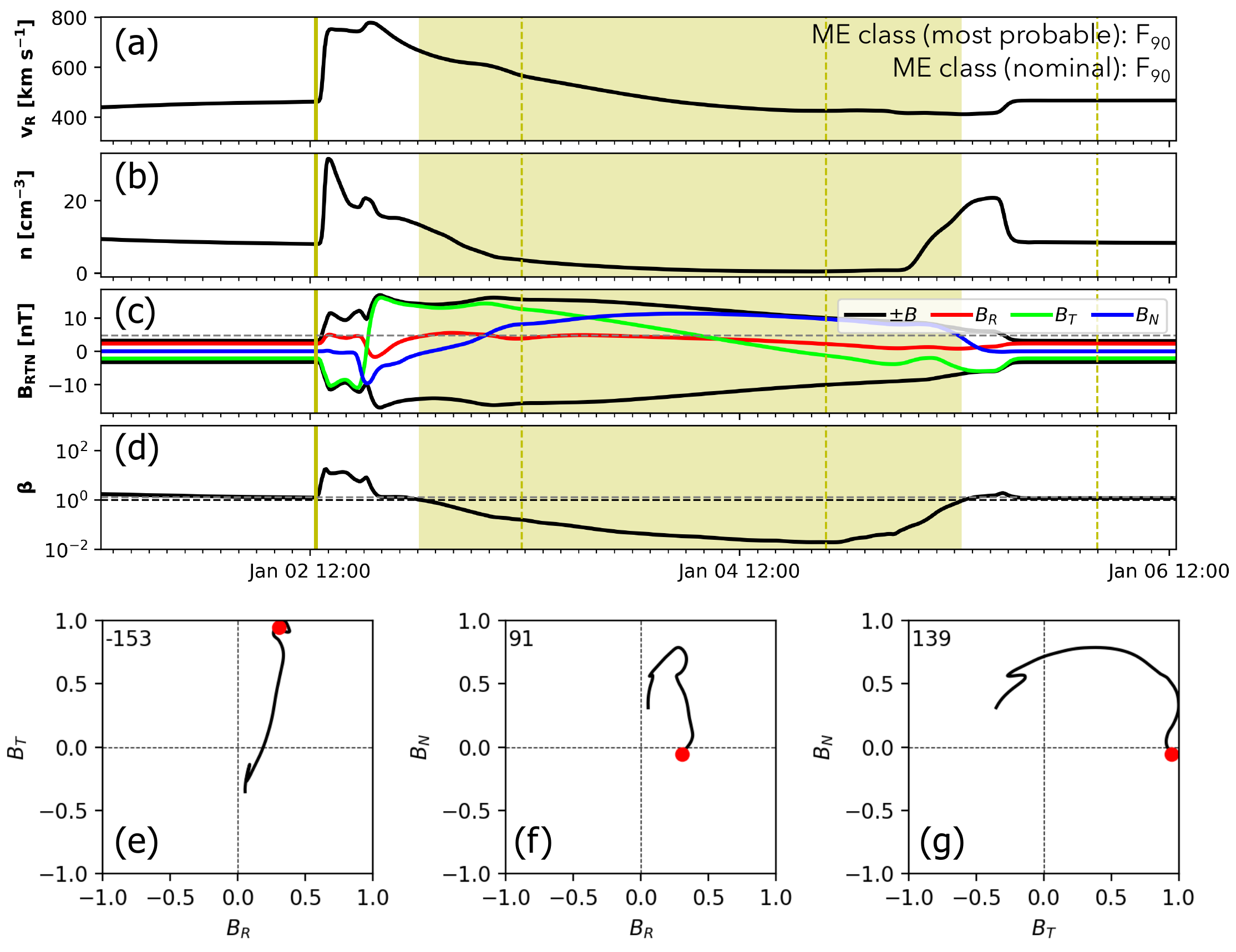}}
\caption{
Example of ME identification and classification from run~A at $r=1$~au, $\theta=5^\circ$, $\phi=20^\circ$ ($F_{90}$ class).
The panels show the same quantities as in Figure~\ref{fig:F270}.
} 
\label{fig:F90}
\end{figure*}

\begin{figure*}
\centering
{\includegraphics[width=0.7\hsize]{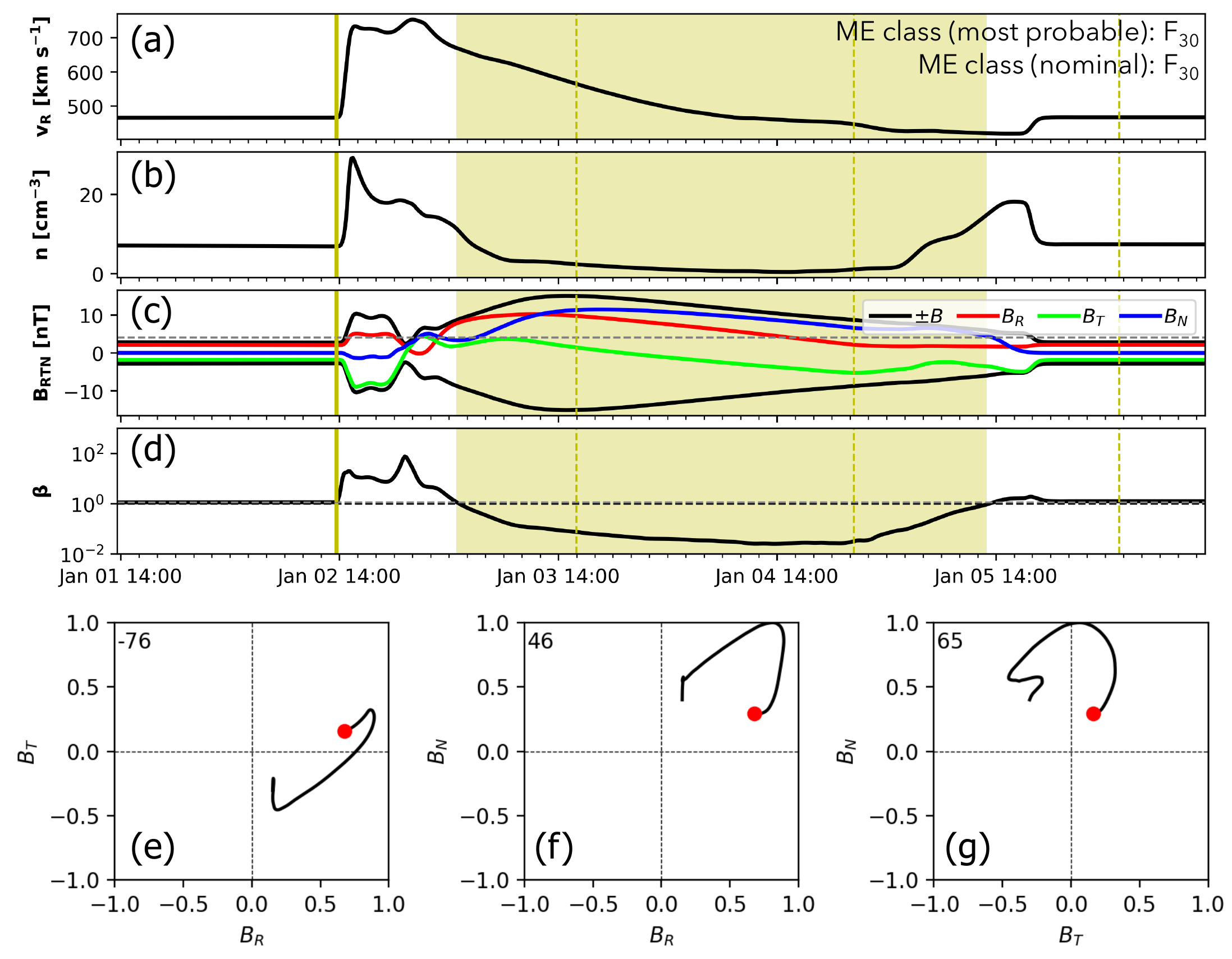}}
\caption{
Example of ME identification and classification from run~A at $r=1$~au, $\theta=15^\circ$, $\phi=20^\circ$ ($F_{30}$ class).
The panels show the same quantities as in Figure~\ref{fig:F270}.
} 
\label{fig:F30}
\end{figure*}

\begin{figure*}
\centering
{\includegraphics[width=0.7\hsize]{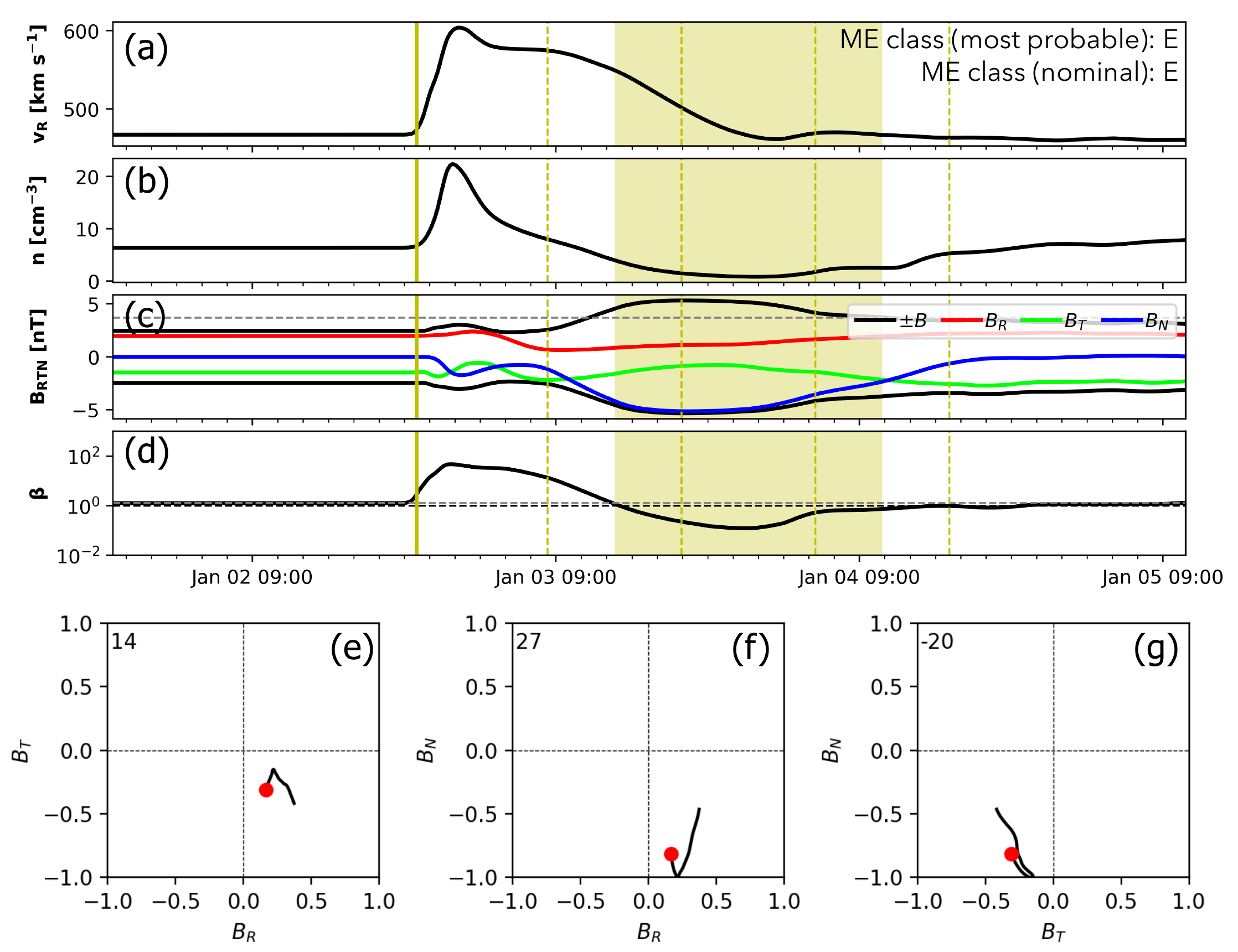}}
\caption{
Example of ME identification and classification from run~A at $r=1$~au, $\theta=30^\circ$, $\phi=-45^\circ$ ($E$ class).
The panels show the same quantities as in Figure~\ref{fig:F270}.
} 
\label{fig:E}
\end{figure*}

\begin{figure*}
\centering
{\includegraphics[width=0.7\hsize]{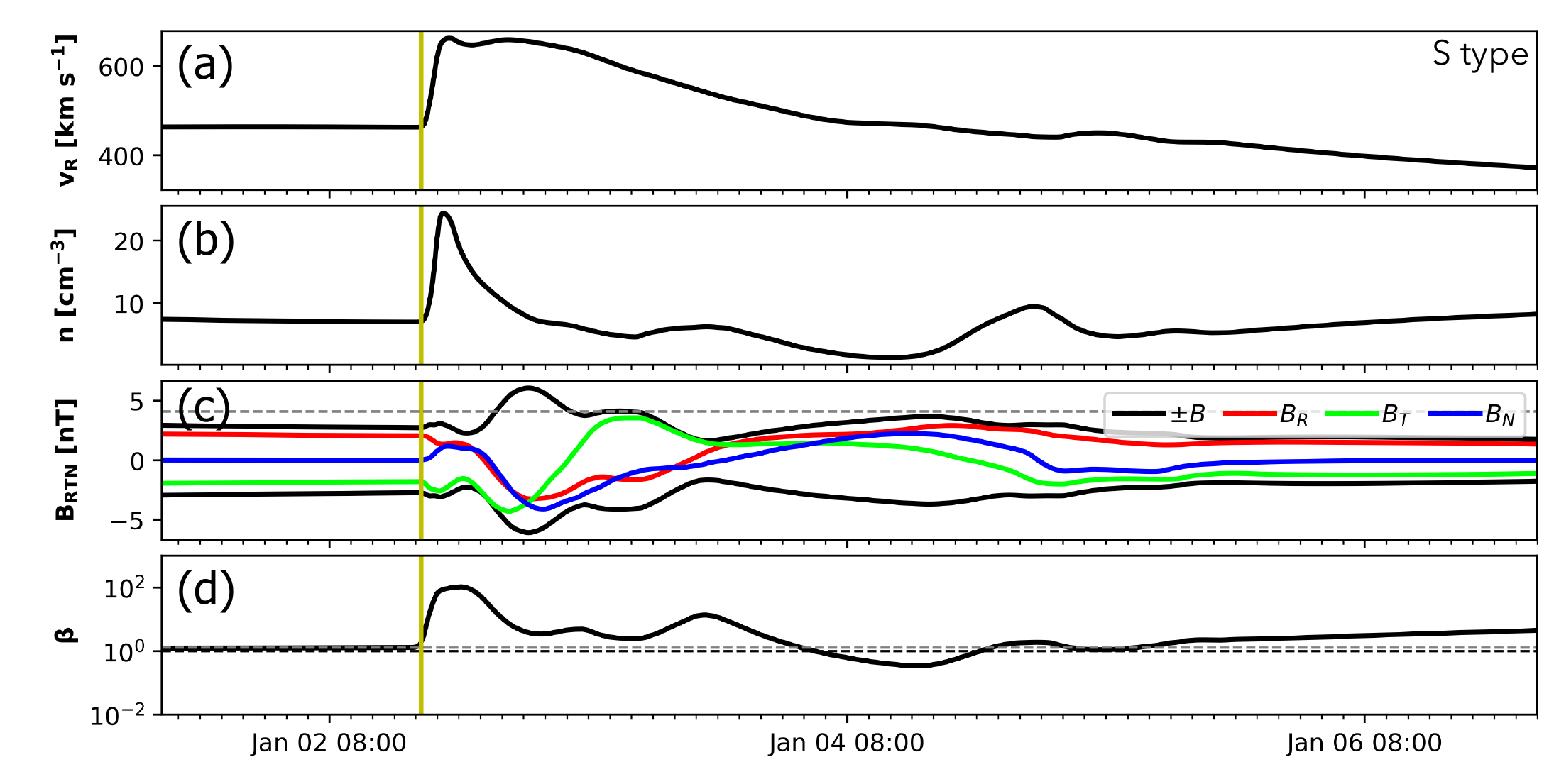}}
\caption{
Example of ME identification and classification from run~A at $r=1$~au, $\theta=0^\circ$, $\phi=-40^\circ$ ($S$ class).
(a), (b), (c), (d): time series for the radial speed, number density, magnetic field components, and plasma $\beta$.
The shock-like perturbation is marked by the continuous yellow line.
} 
\label{fig:S}
\end{figure*}

Examples of the various ME and shock signatures identified at different virtual spacecraft located at 1~au in run~A are provided in Figures~\ref{fig:F180} ($F_{180}$ ME class), \ref{fig:F90} ($F_{90}$ ME class), \ref{fig:F30} ($F_{30}$ ME class), \ref{fig:E} ($E$ ME class), and \ref{fig:S} ($S$ class).

\bibliography{Refs}{}

\begin{thebibliography}{}
\expandafter\ifx\csname natexlab\endcsname\relax\def\natexlab#1{#1}\fi
\providecommand{\url}[1]{\href{#1}{#1}}
\providecommand{\dodoi}[1]{doi:~\href{http://doi.org/#1}{\nolinkurl{#1}}}
\providecommand{\doeprint}[1]{\href{http://ascl.net/#1}{\nolinkurl{http://ascl.net/#1}}}
\providecommand{\doarXiv}[1]{\href{https://arxiv.org/abs/#1}{\nolinkurl{https://arxiv.org/abs/#1}}}

\bibitem[{{Al-Haddad} {et~al.}(2019){Al-Haddad}, {Poedts}, {Roussev},
  {Farrugia}, {Yu}, \& {Lugaz}}]{Al-Haddad2019}
{Al-Haddad}, N., {Poedts}, S., {Roussev}, I., {et~al.} 2019, \apj, 870, 100,
  \dodoi{10.3847/1538-4357/aaf38d}

\bibitem[{{Al-Haddad} {et~al.}(2013){Al-Haddad}, {Nieves-Chinchilla}, {Savani},
  {M{\"o}stl}, {Marubashi}, {Hidalgo}, {Roussev}, {Poedts}, \&
  {Farrugia}}]{Al-Haddad2013}
{Al-Haddad}, N., {Nieves-Chinchilla}, T., {Savani}, N.~P., {et~al.} 2013,
  \solphys, 284, 129, \dodoi{10.1007/s11207-013-0244-5}

\bibitem[{{Bothmer} \& {Schwenn}(1998)}]{Bothmer1998}
{Bothmer}, V., \& {Schwenn}, R. 1998, Annales Geophysicae, 16, 1,
  \dodoi{10.1007/s00585-997-0001-x}

\bibitem[{{Burlaga} {et~al.}(1981){Burlaga}, {Sittler}, {Mariani}, \&
  {Schwenn}}]{Burlaga1981}
{Burlaga}, L., {Sittler}, E., {Mariani}, F., \& {Schwenn}, R. 1981, \jgr, 86,
  6673, \dodoi{10.1029/JA086iA08p06673}

\bibitem[{{Cranmer} {et~al.}(2017){Cranmer}, {Gibson}, \&
  {Riley}}]{Cranmer2017}
{Cranmer}, S.~R., {Gibson}, S.~E., \& {Riley}, P. 2017, \ssr, 212, 1345,
  \dodoi{10.1007/s11214-017-0416-y}

\bibitem[{{Davies} {et~al.}(2020{\natexlab{a}}){Davies}, {Forsyth}, {Good}, \&
  {Kilpua}}]{Davies2020}
{Davies}, E.~E., {Forsyth}, R.~J., {Good}, S.~W., \& {Kilpua}, E. K.~J.
  2020{\natexlab{a}}, \solphys, 295, 157, \dodoi{10.1007/s11207-020-01714-z}

\bibitem[{{Davies} {et~al.}(2020{\natexlab{b}}){Davies}, {M{\"o}stl}, {Owens},
  {Weiss}, {Amerstorfer}, {Hinterreiter}, {Bauer}, {Bailey}, {Reiss},
  {Forsyth}, {Horbury}, {O'Brien}, {Evans}, {Angelini}, {Heyner}, {Richter},
  {Auster}, {Magnes}, {Baumjohann}, {Fischer}, {Barnes}, {Davies}, \&
  {Harrison}}]{Davies2021}
{Davies}, E.~E., {M{\"o}stl}, C., {Owens}, M.~J., {et~al.} 2020{\natexlab{b}},
  arXiv e-prints, arXiv:2012.07456.
\newblock \doarXiv{2012.07456}

\bibitem[{{Dungey}(1961)}]{Dungey1961}
{Dungey}, J.~W. 1961, \prl, 6, 47, \dodoi{10.1103/PhysRevLett.6.47}

\bibitem[{{Farrugia} {et~al.}(1993){Farrugia}, {Burlaga}, {Osherovich},
  {Richardson}, {Freeman}, {Lepping}, \& {Lazarus}}]{Farrugia1993}
{Farrugia}, C.~J., {Burlaga}, L.~F., {Osherovich}, V.~A., {et~al.} 1993, \jgr,
  98, 7621, \dodoi{10.1029/92JA02349}

\bibitem[{{Good} \& {Forsyth}(2016)}]{Good2016}
{Good}, S.~W., \& {Forsyth}, R.~J. 2016, \solphys, 291, 239,
  \dodoi{10.1007/s11207-015-0828-3}

\bibitem[{{Good} {et~al.}(2018){Good}, {Forsyth}, {Eastwood}, \&
  {M{\"o}stl}}]{Good2018}
{Good}, S.~W., {Forsyth}, R.~J., {Eastwood}, J.~P., \& {M{\"o}stl}, C. 2018,
  \solphys, 293, 52, \dodoi{10.1007/s11207-018-1264-y}

\bibitem[{{Good} {et~al.}(2015){Good}, {Forsyth}, {Raines}, {Gershman},
  {Slavin}, \& {Zurbuchen}}]{Good2015}
{Good}, S.~W., {Forsyth}, R.~J., {Raines}, J.~M., {et~al.} 2015, \apj, 807,
  177, \dodoi{10.1088/0004-637X/807/2/177}

\bibitem[{{Good} {et~al.}(2019){Good}, {Kilpua}, {LaMoury}, {Forsyth},
  {Eastwood}, \& {M{\"o}stl}}]{Good2019}
{Good}, S.~W., {Kilpua}, E.~K.~J., {LaMoury}, A.~T., {et~al.} 2019, J. Geophys.
  Res. (Space Phys.), 124, 4960, \dodoi{10.1029/2019JA026475}

\bibitem[{{Jacobs} {et~al.}(2005){Jacobs}, {Poedts}, {Van der Holst}, \&
  {Chan{\'e}}}]{Jacobs2005}
{Jacobs}, C., {Poedts}, S., {Van der Holst}, B., \& {Chan{\'e}}, E. 2005, \aap,
  430, 1099, \dodoi{10.1051/0004-6361:20041676}

\bibitem[{{Janvier} {et~al.}(2019){Janvier}, {Winslow}, {Good}, {Bonhomme},
  {D{\'e}moulin}, {Dasso}, {M{\"o}stl}, {Lugaz}, {Amerstorfer}, {Soubri{\'e}},
  \& {Boakes}}]{Janvier2019}
{Janvier}, M., {Winslow}, R.~M., {Good}, S., {et~al.} 2019, J. Geophys. Res.
  (Space Phys.), 124, 812, \dodoi{10.1029/2018JA025949}

\bibitem[{{Jian} {et~al.}(2018){Jian}, {Russell}, {Luhmann}, \&
  {Galvin}}]{Jian2018}
{Jian}, L.~K., {Russell}, C.~T., {Luhmann}, J.~G., \& {Galvin}, A.~B. 2018,
  \apj, 855, 114, \dodoi{10.3847/1538-4357/aab189}

\bibitem[{{Kilpua} {et~al.}(2017{\natexlab{a}}){Kilpua}, {Koskinen}, \&
  {Pulkkinen}}]{Kilpua2017}
{Kilpua}, E., {Koskinen}, H. E.~J., \& {Pulkkinen}, T.~I. 2017{\natexlab{a}},
  Liv. Rev. Sol. Phys., 14, 5, \dodoi{10.1007/s41116-017-0009-6}

\bibitem[{{Kilpua} {et~al.}(2017{\natexlab{b}}){Kilpua}, {Balogh}, {von
  Steiger}, \& {Liu}}]{Kilpua2017b}
{Kilpua}, E.~K.~J., {Balogh}, A., {von Steiger}, R., \& {Liu}, Y.~D.
  2017{\natexlab{b}}, \ssr, 212, 1271, \dodoi{10.1007/s11214-017-0411-3}

\bibitem[{{Kilpua} {et~al.}(2015){Kilpua}, {Lumme}, {Andreeova}, {Isavnin}, \&
  {Koskinen}}]{Kilpua2015}
{Kilpua}, E.~K.~J., {Lumme}, E., {Andreeova}, K., {Isavnin}, A., \& {Koskinen},
  H.~E.~J. 2015, J. Geophys. Res. (Space Phys.), 120, 4112,
  \dodoi{10.1002/2015JA021138}

\bibitem[{{Lee} {et~al.}(2017){Lee}, {Hara}, {Halekas}, {Thiemann},
  {Chamberlin}, {Eparvier}, {Lillis}, {Larson}, {Dunn}, {Espley}, {Gruesbeck},
  {Curry}, {Luhmann}, \& {Jakosky}}]{Lee2017}
{Lee}, C.~O., {Hara}, T., {Halekas}, J.~S., {et~al.} 2017, J. Geophys. Res.
  (Space Phys.), 122, 2768, \dodoi{10.1002/2016JA023495}

\bibitem[{{Lugaz} {et~al.}(2011){Lugaz}, {Downs}, {Shibata}, {Roussev}, {Asai},
  \& {Gombosi}}]{Lugaz2011}
{Lugaz}, N., {Downs}, C., {Shibata}, K., {et~al.} 2011, \apj, 738, 127,
  \dodoi{10.1088/0004-637X/738/2/127}

\bibitem[{{Lugaz} {et~al.}(2018){Lugaz}, {Farrugia}, {Winslow}, {Al-Haddad},
  {Galvin}, {Nieves-Chinchilla}, {Lee}, \& {Janvier}}]{Lugaz2018}
{Lugaz}, N., {Farrugia}, C.~J., {Winslow}, R.~M., {et~al.} 2018, \apjl, 864,
  L7, \dodoi{10.3847/2041-8213/aad9f4}

\bibitem[{{Lugaz} {et~al.}(2020{\natexlab{a}}){Lugaz}, {Salman}, {Winslow},
  {Al-Haddad}, {Farrugia}, {Zhuang}, \& {Galvin}}]{Lugaz2020b}
{Lugaz}, N., {Salman}, T.~M., {Winslow}, R.~M., {et~al.} 2020{\natexlab{a}},
  \apj, 899, 119, \dodoi{10.3847/1538-4357/aba26b}

\bibitem[{{Lugaz} {et~al.}(2017){Lugaz}, {Temmer}, {Wang}, \&
  {Farrugia}}]{Lugaz2017}
{Lugaz}, N., {Temmer}, M., {Wang}, Y., \& {Farrugia}, C.~J. 2017, \solphys,
  292, 64, \dodoi{10.1007/s11207-017-1091-6}

\bibitem[{{Lugaz} {et~al.}(2020{\natexlab{b}}){Lugaz}, {Winslow}, \&
  {Farrugia}}]{Lugaz2020}
{Lugaz}, N., {Winslow}, R.~M., \& {Farrugia}, C.~J. 2020{\natexlab{b}}, J.
  Geophys. Res. (Space Phys.), 125, e27213, \dodoi{10.1029/2019JA027213}

\bibitem[{{Manchester} {et~al.}(2017){Manchester}, {Kilpua}, {Liu}, {Lugaz},
  {Riley}, {T{\"o}r{\"o}k}, \& {Vr{\v s}nak}}]{Manchester2017}
{Manchester}, W., {Kilpua}, E.~K.~J., {Liu}, Y.~D., {et~al.} 2017, Space
  Science Reviews, 212, 1159, \dodoi{10.1007/s11214-017-0394-0}

\bibitem[{{Manchester} {et~al.}(2004){Manchester}, {Gombosi}, {Roussev}, {de
  Zeeuw}, {Sokolov}, {Powell}, {T{\'o}th}, \& {Opher}}]{Manchester2004}
{Manchester}, W.~B., {Gombosi}, T.~I., {Roussev}, I., {et~al.} 2004, Journal of
  Geophysical Research (Space Physics), 109, A01102,
  \dodoi{10.1029/2002JA009672}

\bibitem[{{McComas} {et~al.}(2008){McComas}, {Ebert}, {Elliott}, {Goldstein},
  {Gosling}, {Schwadron}, \& {Skoug}}]{McComas2008}
{McComas}, D.~J., {Ebert}, R.~W., {Elliott}, H.~A., {et~al.} 2008, \grl, 35,
  L18103, \dodoi{10.1029/2008GL034896}

\bibitem[{{Nieves-Chinchilla} {et~al.}(2019){Nieves-Chinchilla}, {Jian},
  {Balmaceda}, {Vourlidas}, {dos Santos}, \& {Szabo}}]{Nieves2019}
{Nieves-Chinchilla}, T., {Jian}, L.~K., {Balmaceda}, L., {et~al.} 2019,
  \solphys, 294, 89, \dodoi{10.1007/s11207-019-1477-8}

\bibitem[{{Odstr{\v{c}}il} {et~al.}(1996){Odstr{\v{c}}il}, {Dryer}, \&
  {Smith}}]{Odstrcil1996}
{Odstr{\v{c}}il}, D., {Dryer}, M., \& {Smith}, Z. 1996, \jgr, 101, 19973,
  \dodoi{10.1029/96JA00479}

\bibitem[{{Odstr{\v{c}}il} \& {Pizzo}(1999)}]{Odstrcil1999a}
{Odstr{\v{c}}il}, D., \& {Pizzo}, V.~J. 1999, \jgr, 104, 28225,
  \dodoi{10.1029/1999JA900319}

\bibitem[{{Owens} {et~al.}(2017){Owens}, {Lockwood}, \& {Barnard}}]{Owens2017}
{Owens}, M.~J., {Lockwood}, M., \& {Barnard}, L.~A. 2017, Scientific Reports,
  7, 4152, \dodoi{10.1038/s41598-017-04546-3}

\bibitem[{{Pal} {et~al.}(2018){Pal}, {Nandy}, {Srivastava}, {Gopalswamy}, \&
  {Panda}}]{Pal2018}
{Pal}, S., {Nandy}, D., {Srivastava}, N., {Gopalswamy}, N., \& {Panda}, S.
  2018, \apj, 865, 4, \dodoi{10.3847/1538-4357/aada10}

\bibitem[{{Pomoell} \& {Poedts}(2018)}]{Pomoell2018}
{Pomoell}, J., \& {Poedts}, S. 2018, J. Space Weather Space Clim., 8, A35,
  \dodoi{10.1051/swsc/2018020}

\bibitem[{{Richardson} \& {Cane}(2010)}]{Richardson2010}
{Richardson}, I.~G., \& {Cane}, H.~V. 2010, \solphys, 264, 189,
  \dodoi{10.1007/s11207-010-9568-6}

\bibitem[{{Riley} {et~al.}(2004){Riley}, {Linker}, {Lionello}, {Miki{\'c}},
  {Odstrcil}, {Hidalgo}, {Cid}, {Hu}, {Lepping}, {Lynch}, \&
  {Rees}}]{Riley2004}
{Riley}, P., {Linker}, J.~A., {Lionello}, R., {et~al.} 2004, J. Atmos.
  Sol.-Terr. Phys, 66, 1321, \dodoi{10.1016/j.jastp.2004.03.019}

\bibitem[{{Rouillard} {et~al.}(2010){Rouillard}, {Lavraud}, {Sheeley},
  {Davies}, {Burlaga}, {Savani}, {Jacquey}, \& {Forsyth}}]{Rouillard2010}
{Rouillard}, A.~P., {Lavraud}, B., {Sheeley}, N.~R., {et~al.} 2010, \apj, 719,
  1385, \dodoi{10.1088/0004-637X/719/2/1385}

\bibitem[{{Salman} {et~al.}(2020){Salman}, {Winslow}, \& {Lugaz}}]{Salman2020}
{Salman}, T.~M., {Winslow}, R.~M., \& {Lugaz}, N. 2020, J. Geophys. Res. (Space
  Phys.), 125, e27084, \dodoi{10.1029/2019JA027084}

\bibitem[{{Scolini} {et~al.}(2021){Scolini}, {Dasso}, {Rodriguez}, {Zhukov}, \&
  {Poedts}}]{Scolini2020A&A}
{Scolini}, C., {Dasso}, S., {Rodriguez}, L., {Zhukov}, A.~N., \& {Poedts}, S.
  2021, A\&A (in press)

\bibitem[{{Scolini} {et~al.}(2019){Scolini}, {Rodriguez}, {Mierla}, {Pomoell},
  \& {Poedts}}]{Scolini2019}
{Scolini}, C., {Rodriguez}, L., {Mierla}, M., {Pomoell}, J., \& {Poedts}, S.
  2019, \aap, 626, A122, \dodoi{10.1051/0004-6361/201935053}

\bibitem[{{Sharma} \& {Cid}(2020)}]{Sharma2020}
{Sharma}, R., \& {Cid}, C. 2020, \aap, 642, A233,
  \dodoi{10.1051/0004-6361/202038927}

\bibitem[{{T{\"o}r{\"o}k} {et~al.}(2018){T{\"o}r{\"o}k}, {Downs}, {Linker},
  {Lionello}, {Titov}, {Miki{\'c}}, {Riley}, {Caplan}, \& {Wijaya}}]{Torok2018}
{T{\"o}r{\"o}k}, T., {Downs}, C., {Linker}, J.~A., {et~al.} 2018, \apj, 856,
  75, \dodoi{10.3847/1538-4357/aab36d}

\bibitem[{{Tsurutani} {et~al.}(1988){Tsurutani}, {Gonzalez}, {Tang}, {Akasofu},
  \& {Smith}}]{Tsurutani1988}
{Tsurutani}, B.~T., {Gonzalez}, W.~D., {Tang}, F., {Akasofu}, S.~I., \&
  {Smith}, E.~J. 1988, \jgr, 93, 8519, \dodoi{10.1029/JA093iA08p08519}

\bibitem[{{Verbeke} {et~al.}(2019){Verbeke}, {Pomoell}, \&
  {Poedts}}]{Verbeke2019}
{Verbeke}, C., {Pomoell}, J., \& {Poedts}, S. 2019, \aap, 627, A111,
  \dodoi{10.1051/0004-6361/201834702}

\bibitem[{{Vorotnikov} {et~al.}(2008){Vorotnikov}, {Smith}, {Hu}, {Szabo},
  {Skoug}, \& {Cohen}}]{Vorotnikov2008}
{Vorotnikov}, V.~S., {Smith}, C.~W., {Hu}, Q., {et~al.} 2008, Space Weather, 6,
  03002, \dodoi{10.1029/2007SW000358}

\bibitem[{{Vourlidas} {et~al.}(2013){Vourlidas}, {Lynch}, {Howard}, \&
  {Li}}]{Vourlidas2013}
{Vourlidas}, A., {Lynch}, B.~J., {Howard}, R.~A., \& {Li}, Y. 2013, \solphys,
  284, 179, \dodoi{10.1007/s11207-012-0084-8}

\bibitem[{{Vr{\v{s}}nak} {et~al.}(2019){Vr{\v{s}}nak}, {Amerstorfer},
  {Dumbovi{\'c}}, {Leitner}, {Veronig}, {Temmer}, {M{\"o}stl}, {Amerstorfer},
  {Farrugia}, \& {Galvin}}]{Vrsnak2019}
{Vr{\v{s}}nak}, B., {Amerstorfer}, T., {Dumbovi{\'c}}, M., {et~al.} 2019, \apj,
  877, 77, \dodoi{10.3847/1538-4357/ab190a}

\bibitem[{{Wang} {et~al.}(2004){Wang}, {Shen}, {Wang}, \& {Ye}}]{Wang2004}
{Wang}, Y., {Shen}, C., {Wang}, S., \& {Ye}, P. 2004, \solphys, 222, 329,
  \dodoi{10.1023/B:SOLA.0000043576.21942.aa}

\bibitem[{{Webb} \& {Howard}(2012)}]{Webb2012}
{Webb}, D.~F., \& {Howard}, T.~A. 2012, Liv. Rev. Sol. Phys., 9, 3,
  \dodoi{10.12942/lrsp-2012-3}

\bibitem[{{Winslow} {et~al.}(2015){Winslow}, {Lugaz}, {Philpott}, {Schwadron},
  {Farrugia}, {Anderson}, \& {Smith}}]{Winslow2015}
{Winslow}, R.~M., {Lugaz}, N., {Philpott}, L.~C., {et~al.} 2015, J. Geophys.
  Res. (Space Phys.), 120, 6101, \dodoi{10.1002/2015JA021200}

\bibitem[{{Winslow} {et~al.}(2021){Winslow}, {Scolini}, {Lugaz}, \&
  {Galvin}}]{Winslow2021}
{Winslow}, R.~M., {Scolini}, C., {Lugaz}, N., \& {Galvin}, A.~B. 2021, ApJ
  (accepted)

\bibitem[{{Winslow} {et~al.}(2016){Winslow}, {Lugaz}, {Schwadron}, {Farrugia},
  {Yu}, {Raines}, {Mays}, {Galvin}, \& {Zurbuchen}}]{Winslow2016}
{Winslow}, R.~M., {Lugaz}, N., {Schwadron}, N.~A., {et~al.} 2016, J. Geophys.
  Res. (Space Phys.), 121, 6092, \dodoi{10.1002/2015JA022307}

\bibitem[{{Winslow} {et~al.}(2020){Winslow}, {Lugaz}, {Philpott}, {Farrugia},
  {Johnson}, {Anderson}, {Paty}, {Schwadron}, \& {Asad}}]{Winslow2020}
{Winslow}, R.~M., {Lugaz}, N., {Philpott}, L., {et~al.} 2020, \apj, 889, 184,
  \dodoi{10.3847/1538-4357/ab6170}

\bibitem[{{Zhang} {et~al.}(2007){Zhang}, {Richardson}, {Webb}, {Gopalswamy},
  {Huttunen}, {Kasper}, {Nitta}, {Poomvises}, {Thompson}, {Wu}, {Yashiro}, \&
  {Zhukov}}]{Zhang2007}
{Zhang}, J., {Richardson}, I.~G., {Webb}, D.~F., {et~al.} 2007, J. Geophys.
  Res. (Space Phys.), 112, A10102, \dodoi{10.1029/2007JA012321}

\end{thebibliography}
\bibliographystyle{aasjournal}



\end{document}